\documentclass[12pt]{article}

\usepackage[pagebackref=false,colorlinks=true,citecolor=blue,filecolor=blue,%
linkcolor=blue,anchorcolor=blue,urlcolor=blue]{hyperref}

\usepackage{amsmath}
\usepackage{slashed}
\usepackage[dvips]{graphicx}

\newcommand{\nn}{\nonumber}
\newcommand{\req}[1]{(\ref{#1})}
\newcommand{\be}{\begin{equation}}
\newcommand{\ee}{\end{equation}}
\newcommand{\ba}{\begin{eqnarray}}
\newcommand{\ea}{\end{eqnarray}}
\newcommand{\ci}[1]{\cite{#1}}
\newcommand{\tw}{\textwidth}                          

\newcommand{\da}{{DA}}

\def\={\,=\,}

\def\vk{{\bf k}_{\perp}}

\newcommand{\LQCD}{\Lambda_{\rm{QCD}}}
\def\als{\alpha_s}

\def\mev{\,{\rm MeV}}
\def\gev{\,{\rm GeV}}

\def\muR{\mu_R}

\def\Li{\relax\ifmmode{\text{Li}_{2}}\else{Li$_2${ }}\fi}

\def\eps{\epsilon}
                 
\def\sh{\hat{s}}
\def\uh{\hat{u}}
\def\th{\hat{t}}

\def\taub{\bar{\tau}}

\newcommand{\bm}{m}

\newcommand{\phiPp}{\phi_{\pi p}}
\newcommand{\phiThreeP}{\phi_{3 \pi}}

\newcommand{\PsiPp}{\Psi_{\pi p}}
\newcommand{\PsiThreeP}{\Psi_{3 \pi}}

\newcommand{\aPp}{a_{\pi p}}
\newcommand{\aThreeP}{a_{3 \pi}}

\newcommand{\OmegaThreeP}{\Omega_{3 \pi}}
\newcommand{\hOmegaThreeP}{\hat{\Omega}_{3 \pi}}

\begin{document}
\thispagestyle{empty}
\begin{flushright}
RBI-ThPhys-2023-49\\
WU-B 23-01\\
\today
\end{flushright}

\begin{center}

{\Large\bf Twist-3 contribution to deeply virtual  \\[0.3em]
electroproduction of pions} \\
\vskip 15mm
G.\ Duplan\v{c}i\'{c}\\[1em]
{\small {\it Theoretical Physics Division, Rudjer Bo\v{s}kovi\'{c} Institute, 
HR-10000 Zagreb, Croatia}}\\[1em]
P.\ Kroll \\[1em]
{\small {\it Fachbereich Physik, Universit\"at Wuppertal, D-42097 Wuppertal,
Germany}}\\[1em]
K.\  Passek-K.\\[1em]
{\small {\it Theoretical Physics Division, Rudjer Bo\v{s}kovi\'{c} Institute, 
HR-10000 Zagreb, Croatia}}\\[1em]
L. Szymanowski\\[1em] National Centre for Nuclear Research (NCBJ), 02-093 Warsaw, Poland\\
\end{center}

\begin{abstract}
The twist-3 contribution, consisting of twist-2 transversity generalized parton
distributions (GPDs) and a twist-3 meson wave function, to deeply virtual pion
electroproduction is discussed.
The twist-3 meson wave function includes both the $q\bar{q}$ and the $q\bar{q}g$
Fock components.
  Two methods to regularize the end-point singularities are introduced - quark transverse momenta
  and a gluon mass. Using existing GPD parameterizations the transverse and the transverse-transverse
  interference cross sections for $\pi^0$ production are calculated and compared to experimental data.
\end{abstract}

\section{Introduction}
\label{sec:introduction}

It has been shown \ci{collins96} that in the generalized Bjorken regime of large photon virtuality ($Q^2$)
and large invariant mass of the hadrons in the final state ($W$) but fixed Bjorken-$x$  ($x_B$) and
squared momentum transfer ($t$) much smaller than $Q^2$, the amplitudes for exclusive meson
electroproduction factorize into generalized parton distributions (GPDs) and perturbatively
calculable subprocess amplitudes. 
The contributions to cross sections from longitudinally polarized photons dominate
in that regime, while those from transversely polarized photons are suppressed by $1/Q^2$, 
leaving aside logarithmic $Q^2$-dependencies. 
However, it is important to realize that it
is theoretically unknown how large $Q^2$ and $W$ must be 
for the factorization concept to hold. 
Thus, from extensive experimental and theoretical investigations it turned out that for
deeply virtual electroproduction of pseudoscalar mesons (DVMP) for which experimental data
are available for $Q^2<10\,\gev^2$, the longitudinal cross section is smaller than the
transverse one, leaving aside the meson-pole contributions.
This is most obvious from the Rosenbluth measurement of the separated cross sections for $\pi^0$
production carried out by the Hall A collaboration at Jefferson Lab \ci{defurne,mazouz}: 
$d\sigma_L\ll d\sigma_T$ is in fact compatible with zero at $Q^2$ of about $2\,\gev^2$ and $x_B\simeq 0.36$
within experimental errors. Large contributions
from transversally polarized photons are as well seen by HERMES in asymmetries for $\pi^+$ production measured
with a transversally polarized proton target \ci{hermes10}. 
The large absolute value of the transverse-transverse interference cross section 
for $\pi^0$ electroproduction \ci{CLAS14} 
also signals strong contributions from transversally polarized photons.

In order to achieve an understanding of the experimental data the transverse amplitudes have been modeled
in \ci{GK5} by twist-2 transversity (or helicity-flip) GPDs in combination with a twist-3 meson
wave function in Wandzura-Wilczek (WW) approximation, i.e.\ by ignoring the 3-body ($q\bar{q}g$) 
Fock component of the meson.
It is to be stressed that the factorization proof given in \ci{collins96}, does not apply to the transverse
amplitudes. Namely, they suffer from  an end-point singularity which has been regularized in \ci{GK5}
by allowing for quark transverse momenta in the meson. The emission and reabsorption of quarks
from the nucleon are still treated collinearly to the nucleon momenta. By the quark transverse momenta in the
meson wave function one effectively takes into account the transverse size of the meson which is neglected in
the usual collinear approach~\footnote{
          The role of the meson's transverse size in diffractive electroproduction of vector mesons has been
          investigated in \ci{frankfurt}.}.
This so-called modified perturbative approach (MPA) describes the data on electroproduction of
pseudoscalar mesons \ci{GK5,GK6,PK-kaon} rather well. 
Similar ideas have also been discussed in \ci{goldstein}.
We remark that the twist-3 effect advocated for in \ci{GK5}
also occurs in exclusive electroproduction of longitudinally polarized vector mesons 
but there it is a little effect visible only in some of the spin-density 
matrix elements \ci{GK7}.  That is in agreement with
experiments \ci{hermes-SDME,compass-SDME}. 
The twist-3 effect which we discuss here, despite similarities,
differs from the one advocated for by Anikin and Teryaev \ci{anikin02} 
for electroproduction of transversally polarized $\rho$ mesons. 
Their twist-3 effect  consists of the usual twist-2 helicity non-flip GPDs in
combination with a twist-3 vector-meson wave function. 

In our recent investigation of wide-angle pion electroproduction \cite{KPK21}, we went
beyond the WW approximation utilized in \cite{GK5} 
and computed the contribution from 3-body ($q\bar{q}g$) Fock component of the meson.
As a byproduct, we also obtained the 3-body contributions at large $Q^2$ and $t=0$.
The 3-body twist-3 amplitude modifies the WW approximation by an additional 3-body contribution 
and a change of the 2-body twist-3 pion distribution amplitude (DA),
$\phiPp$, generated by the 3-body DA, $\phiThreeP$,  via the equation of motion. 
Here in this work we are going to demonstrate how our twist-3 subprocess 
amplitude can be applied to hard exclusive pion electroproduction.

We will present two methods to deal with the end-point singularities in the transverse
amplitudes. First, we will use the MPA, as in \ci{GK5}. 
In the second approach, we use the usual collinear approach but introduce 
in the gluon propagators a dynamically generated mass,
which reflects the fact that a gluon is a carrier of strong interactions 
most strongly influencing the nonlinear dynamics of the infrared sector of QCD.
As in \cite{KPK21,GK5}, in this work, we do not consider the twist-3 contributions 
from the nucleon, i.e, twist-3 GPDs \cite{Belitsky:2000vx}.
They would lead to further power corrections which we expect to be small.

The plan of the paper is the following: In Sec.\ \ref{sec:tw3} we prepare the twist-3 subprocess amplitudes
calculated in \ci{KPK21} for the use in deeply-virtual processes and present the convolutions of GPDs
and the twist-3 subprocess amplitudes in order to calculate the $s$-channel helicity amplitudes for
electroproduction of pions. Sec.\ \ref{sec:DA}  is devoted to the soft-physics input to the evaluation
of observables for pion electroproduction such as the GPDs, the meson DAs and the respective wave functions.
In the following section, Sec.\ \ref{sec:MPA}, the 
twist-3 contribution is treated within the MPA and the
results compared with experiment. The collinear approach with the gluon mass as regulator of the end-point
singularities is described in Sec.\ \ref{sec:collinear} and compared to experimental data. Finally, in Sec.\
\ref{sec:summary}, we present our conclusions.
\section{The twist-3 subprocess amplitudes}
\label{sec:tw3}
The twist-3 amplitudes for the subprocess, $\gamma^*(\mu) q(\lambda) \to \pi^i q(-\lambda)$, have
been calculated in \ci{KPK21}. Here, $\pi^i$ denotes a pion of charge $i$ and $\lambda$ is the helicity
of the ingoing quark, $\mu$ that one of the virtual photon. We will work in Ji's frame \ci{ji98} in which the
subprocess Mandelstam variables $\sh$ and $\uh$ are related to $Q^2$ by
\be
\sh\=\frac{x-\xi}{2\xi}\,Q^2\,, \qquad \uh\=-\frac{x+\xi}{2\xi}\,Q^2\,,
\label{eq:sh-uh}
\ee
and $\th=t$. Thus, $\sh$ and $\uh$ are of order $Q^2$.
The skewness, $\xi$, is defined by the ratio
\be
\xi\=\frac{(p-p')^+}{(p+p')^+}
\, ,
\ee
where $p^+$ and $p'{}^+$ denote the light-cone plus components of the momenta of the incoming and outgoing
nucleons, respectively. The skewness is related to Bjorken-$x$ by
\be
\xi\=\frac{x_B}{2-x_B}
\, ,
\ee
up to corrections of order $1/Q^2$ (see for instance \ci{braun14}). In Eq.\ \req{eq:sh-uh} $x+\xi$ ($x-\xi$)
is the fraction of the plus component of the average nucleon momenta, $(p+p')/2$, the emitted (reabsorbed) quark
carries.

According to \ci{KPK21} the leading-order 2-body twist-3 subprocess amplitudes 
for the production of a pion of charge
$i$ read in collinear approximation~\footnote{
             We have changed the normalization of the spinors employed in  \ci{KPK21}  to the one
                  used in DVMP. This results in a cancellation of $\sqrt{x^2-\xi^2}$ in the subprocess amplitudes.}
 
\ba
   {\cal H}^{\pi^i, q\bar{q}}_{0-\lambda,\mu\lambda}&=&\sqrt{2}\pi (2\lambda+\mu) \als(\mu_R) {\cal C}^{(ab)}_{\pi^i}
                             f_\pi\mu_\pi \frac{C_F}{N_C}\frac{Q^2}{\xi} \nn\\  
                             &&\times \int_0^1\frac{d\tau}{\tau} \phiPp(\tau)\,
                             \left[\frac{e_a}{(\sh+i\eps)^2}+\frac{e_b}{(\uh+i\eps)^2}\right]\,,
\label{eq:2-body-tw3-sub}
\ea
and the 3-body $C_F$ part:
\ba
{\cal H}^{\pi^i,q\bar{q}g,C_F}_{0-\lambda,\mu\lambda}&=&-\sqrt{2}\pi (2\lambda+\mu) \als(\mu_R) {\cal C}^{(ab)}_{\pi^i}
                           f_{3\pi}\frac{C_F}{N_C}\frac{Q^2}{\xi} \nn\\
                           &&\times
                              \int_0^1\frac{d\tau}{\taub^2} \int_0^{\taub}
                           \frac{d\tau_g}{\taub-\tau_g} \phiThreeP(\tau,\taub-\tau_g,\tau_g)
                           \left[\frac{e_a}{(\sh+i\eps)^2}+\frac{e_b}{(\uh+i\eps)^2}\right]
, 
\nn \\
\label{eq:3-body-tw3-CF}
\ea
with the standard $i\eps$ prescription in the propagators. This is needed in DVMP since the poles $\sh=0$ and $\uh=0$
are reached in contrast to wide-angle pion electroproduction \ci{KPK21}.

The 3-body $C_G$ part is given in \ci{KPK21} in a very compact form,
but more appropriate for our purposes is
to go one step back and use instead the replacement
\be
\frac{e_a}{\sh^2\uh} + \frac{e_b}{\sh\uh^2} \= - \frac1{Q^2}\left(\frac{e_a}{\sh^2} + \frac{e_b}{\uh^2}
                                               + \frac{e_a+e_b}{\sh\uh} \right)\,.
 \label{eq:triple}
\ee
The right hand side of this equation actually corresponds to the true diagrammatic origin of
DVMP contributions%
\footnote{In contrast to the general electroproduction contribution \cite{KPK21}
from which \req{eq:3-body-tw3-CF} and \req{eq:3-body-tw3-CG} were derived in the $t \to 0$ limit, 
for DVMP only the $C_A$ and $C_G$ proportional diagram contributions are different from zero.
Therefore, the $C_F$ proportional part entirely originates from $C_A$ 
and naturally corresponds to a part of the $C_G$ contribution.}.
It is a convenient simplification which may be used in a collinear calculation.
The $C_G$ part then reads
\ba
 {\cal H}^{\pi^i,q\bar{q}g,C_G}_{0-\lambda,\mu\lambda}&=&\sqrt{2}\pi (2\lambda+\mu) \als(\mu_R) {\cal C}^{(ab)}_{\pi^i}
                             f_{3\pi}\frac{C_G}{N_C}\frac{Q^2}{\xi} \nn\\
                &&\times\int_0^1\frac{d\tau}{\taub}
               \int_0^{\taub} \frac{d\tau_g}{\tau_g(\taub-\tau_g)} \phiThreeP(\tau,\taub-\tau_g,\tau_g) \nn\\
                             &&\times \left[ \frac{e_a}{(\sh+i\eps)^2} + \frac{e_b}{(\uh+i\eps)^2}
                               + \frac{e_a+e_b}{(\sh+i\eps)(\uh +i\eps)} \right]\,. 
\label{eq:3-body-tw3-CG}
\ea
                      
The 2-body and 3-body twist-3 \da s, $\phiPp$ and $\phiThreeP$, will be discussed in some detail in Sec.
\ref{sec:DA}. The corresponding decay constants - or normalizations,  since the \da s integrated over the momentum
fractions are normalized to unity - are $f_\pi$ and $f_{3\pi}$, respectively. In the definitions of the DAs we are using
light-cone gauge ($A^+=0$). The momentum fraction the gluon carries
is denoted by $\tau_g$, and $\taub$ is $1-\tau$. The strong coupling, $\als(\muR)$, is evaluated in the one-loop
approximation from $\LQCD=0.181\,\gev$ and four flavors ($n_f=4$). The mass parameter, $\mu_\pi$, is large since it is
given by the square of the pion mass, $m_\pi$, enhanced by the chiral condensate
\be
\mu_\pi\= \frac{m_\pi^2}{m_u+m_d}
\, ,
\label{eq:mupi}
\ee
by means of the divergence of the axial-vector current ($m_u$ and $m_d$ are current quark masses). 
In our numerical studies we take a value of $\mu_\pi(\mu_0)=2\,\gev$ at the initial scale $\mu_0=2\,\gev$. 
As usual,
$C_F=(N_C^2-1)/(2N_C)$ and $C_G=C_F-C_A/2$ are color factors where $N_C$ ($=C_A$) is the number of colors.
The constants $e_a$ and $e_b$ are the quark charges in units of the positron charge, $e_0$. The flavor weight factors
for the various pions are
\be
   {\cal C}^{uu}_{\pi^0}\=-{\cal C}^{dd}_{\pi^0}\=\frac1{\sqrt{2}}\,, \qquad {\cal C}^{ud}_{\pi^+}\={\cal C}^{du}_{\pi^-}\=1\,.
\ee
All other ${\cal C}^{(ab)}_{\pi^i}$ are zero. The summation over the same flavor labels is understood.

Any $t$-dependence of the subprocess amplitude is neglected since, for dimensional reasons, $t$ is to be scaled
by $Q^2$ and according to the premise $-t/Q^2\ll 1$. It has been shown in \ci{KPK21} that the twist-3 contributions
to the longitudinal subprocess amplitudes ($\mu=0$) vanish $\sim \sqrt{-t}$. Similarly suppressed are the twist-2
contributions to the transverse amplitudes. It is also important to note that the twist-3 amplitudes
\req{eq:2-body-tw3-sub},
\req{eq:3-body-tw3-CF} and \req{eq:3-body-tw3-CG} are suppressed by $1/Q$ compared to the asymptotically dominant
twist-2 contributions to the longitudinal subprocess amplitudes.

The helicity amplitudes, ${\cal M}^{\pi^i}_{0\nu',\mu\nu}$, for the process $\gamma^*(\mu) N(\nu)\to \pi^i N'(\nu')$ are
given by convolutions of the transversity GPDs, $H_T$ and $\bar{E}_T$, and the twist-3 subprocess amplitudes \ci{GK6}
(explicit helicities are labeled by their signs or by zero)
\ba
   {\cal M}^{\pi^i}_{0-,++}&=& e_0\sqrt{1-\xi^2}\,\int_{-\xi}^1 dx\, H_T(x,\xi,t) {\cal H}^{\pi^i,tw3}_{0-,++}(x,\xi)\,, \nn\\
   {\cal M}^{\pi^i}_{0+,\pm +}&=& -e_0\frac{\sqrt{-t'}}{4m}\,\int_{-\xi}^1 dx\, \bar{E}_T(x,\xi,t)
                                                 {\cal H}^{\pi^i,tw3}_{0-,++}(x,\xi)\,,  \nn\\
    {\cal M}^{\pi^i}_{0-,-+}&=0\,,  
\label{eq:hel-amp}
\ea
 where
 \be
    {\cal H}^{\pi^i,tw3}_{0-,++}\= {\cal H}^{\pi^i, q\bar{q}}_{0-,++} + {\cal H}^{\pi^i, q\bar{q}g,C_G}_{0-,++}
    + {\cal H}^{\pi^i, q\bar{q}g,C_F}_{0-,++}
\, ,
\ee
and
\be
t'\=t-t_0\,.
\label{eq:tprime}
\ee
The quantity $t_0$ is the  minimal value of $-t$ allowed in the process of interest. It is related to the
skewness by 
\be
t_0\=-4m^2 \frac{\xi^2}{1-\xi^2}\,,
\ee
with $m$ being the mass of the nucleon.
The contributions from other transversity GPDs, as for instance $\widetilde{H}_T$, are neglected. There is no
evidence in the available data for such contributions. 
We also restrict this investigation to valence-quark
GPDs as in \ci{GK5,GK6}.    

Inspection of \req{eq:2-body-tw3-sub} reveals that there is an end-point singularity in ${\cal H}^{\pi^i, q\bar{q}}_{0-,++}$
since $\phiPp(\tau)\to 1$ for $\tau\to 0$ or 1. This singularity requires a regularization for which we
are going to present two methods below: the introduction of quark transverse momenta (Sec. \ref{sec:MPA})
and a gluon mass (Sec. \ref{sec:collinear}). There are no end-point singularities in ${\cal H}^{\pi^i,q\bar{q}g,C_G}$
and ${\cal H}^{\pi^i,q\bar{q}g,C_F}$ since, in contrast to $\phiPp$, the 3-body \da{} $\phiThreeP$ vanishes at the
end-points.

\section{The soft physics input}
\label{sec:DA}
\subsection{GPDs}
\label{sec:GKgpds}
As a starting point for a comparison with experiment we are going to use the GPDs proposed in \ci{GK6,GK7,GK3}.
One should however be aware of possible necessary changes of them in order to fit the experimental data since
the subprocess amplitudes are different now. 
As for the DAs, light-cone gauge is used in the definitions of the GPDs. 

\begin{figure}[t]
\begin{center}
  \includegraphics[width=0.35\tw]{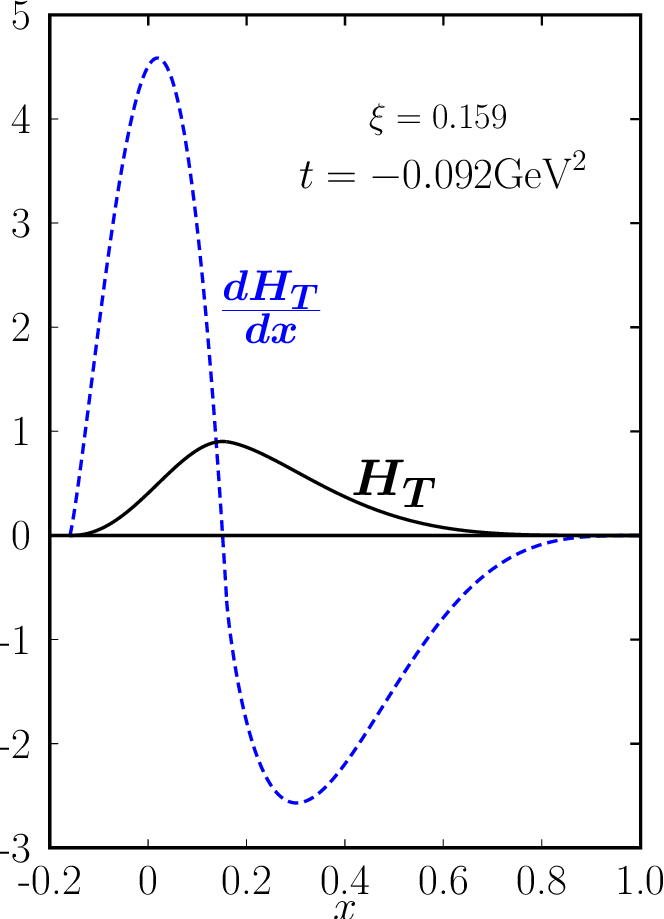}
\end{center}
    \caption{The $\pi^0$ combination of the GPD $H_T$ and its derivative at the initial scale, $\mu_0=2 GeV$.}
    \label{fig:derivative}
\end{figure}

\begin{table*}[t]
  \caption{Parameters of the GPDs at the initial scale $\mu_0=2\,\gev$, see \ci{GK6,GK7}.
       The GPD $\tilde{E}$ is only the non-pole part.
   Parameters for which no value is quoted are fixed by the parton densities.} 
\renewcommand{\arraystretch}{1.2} 
\begin{center}
\begin{tabular}{| c || c  c  c  c c | }
  \hline
  $K(x,\xi,t)$ &  $b$   &  $\alpha(0)$  & $\alpha'$ &   $N$ & $\beta$ \\[0.2em]
  \hline
  $\tilde{H}^{u,d}$ &  0.59  &  0.32   & 0.45 &  -   & - \\[0.2em]
  $\tilde{E}^u$ &  0.9   &  0.48   & 0.45  & 14.0 &  5  \\[0.2em]
  $\tilde{E}^d$ &  0.9   &  0.48   & 0.45  & 4.0  &  5   \\[0.2em]
  $H_T^u$    &  0.3   & -   & 0.45  & 1.1  &  -  \\[0.2em]
  $H_T^d$    &  0.3   & -   & 0.45  & -0.3  & -   \\[0.2em]
  $\bar{E}_T^u$ &  0.77   & -0.10   & 0.45  & 20.91  &   4  \\[0.2em]
  $\bar{E}_T^d$ &  0.5   & -0.10   & 0.45  & 15.46 &     5 \\[0.2em]   
  \hline 
\end{tabular}
\end{center}
\label{tab:GPD-parameters}
\renewcommand{\arraystretch}{1.0}   
\end{table*} 
In \ci{GK6,GK7,GK3} the GPDs are constructed from the zero skewness GPDs. Their products with suitable weight
functions are considered as double distributions from which the skewness dependence of the GPDs is generated
\ci{musatov99}. A zero-skewness GPD for a flavor $a$ ($=u,d$ here) is parameterized as
\be
K_j^a(x,\xi=0,t)\=K_j^a(x,\xi=t=0) \exp{[(b_j^a-\alpha'{}^a_j \ln{x})t]}
\, .
\label{eq:GPD}
\ee
This ansatz is only suitable at small $-t$ since its Mellin moments fall exponentially at large $-t$.
Such a behavior is in conflict with the experimental data on the electromagnetic form factors of the nucleon
which show a power-law decrease~\footnote{\label{footnote4}
     In \ci{DFJK4} a modification of the profile function in \req{eq:GPD} has been proposed: It
     is multiplied by $(1-x)^3$ and a term $Ax(1-x)^2$ added. This new profile function is also suitable
     for large $-t$. The nucleon form factors fall as powers of $t$ for it. This parameterization is supported by
     light-front holographic QCD \ci{teramond}.}.
The forward limit of the GPD $H_T$ is given by the transversity parton density. This forward limit is parameterized as
\be
H_T^a(x,\xi=t=0) \= N^a_{H_T}\sqrt{x} (1-x)[ q^a(x)+\Delta q^a(x)]\,.
\label{eq:HT-parameterization}
\ee
This guarantees that the transversity density respects the Soffer bound. The unpolarized ($q^a(x)$) and
polarized ($\Delta q^a(x)$) densities are taken from Refs.\ \ci{ABM12} and \ci{FSSV09}, respectively.
The forward limits of the $E$-type GPDs are parameterized like the parton densities
\be
K^a_j(x,\xi=t=0)\=N_j^a x^{-\alpha^a_j(0)} (1-x)^{\beta^a_j}
\, .
\ee
The additional parameters are fitted to the meson electroproduction data. 
The various GPD parameters are compiled
in Tab.\ \ref{tab:GPD-parameters}. Occasionally we need the derivative of a GPD with regard to $x$.
In Fig.\ \ref{fig:derivative} we display the GPD $H_T$ and its derivative as an example. One sees
that our twist-2 GPDs as well as their derivatives are continuous at $x=\pm \xi$. 
We remark that in some special models  the derivatives of the GPDs  are non-continuous at $x=\pm \xi$ , see the
discussion in \ci{diehl03}~\footnote{ 
        In some models twist-3 GPDs are even non-continuous at $x =\pm \xi$ \ci{kivel,aslan}.
        In \ci{aslan} is has been conjectured that this may be a general feature of the twist-3 GPDs which
        would lead to problems with factorization. However, as shown in \cite{aslan}, a particular linear combination
        of twist-3 GPDs contributes to deeply virtual Compton scattering for which the discontinuities at $x=\pm \xi$
        cancel. Here, in our work, we do not include the twist-3 GPDs. 
        }.

For the described
parameterization of the zero-skewness GPDs combined with a suitable weight function \ci{musatov99}, the
double-distribution integral can be carried out analytically. The results of this integration are given in \ci{GK3}.
As is well-known the GPDs depend on the scale, see \ci{bel-rad}
and references therein. This evolution effect is taken into account in our numerical studies. 

As we said above the necessity  may turn up to modify the transversity GPDs. However, there are constraints
on these GPDs from lattice QCD: in \ci{LQCD,LQCD2} the first two moments of the transversity
GPDs have been calculated. These results are compared to the moments evaluated from our GPDs in
Tab.\ \ref{tab:moments}. With regard to the uncertainties of the GPDs determined in \ci{GK5,GK6} and those inherent
in the lattice calculations we think there is fair agreement of the first moments although the $d$-quark moments
of the GPD $H_T$ are a bit small.
In \ci{bertone21} it has been pointed out that the GPDs cannot be extracted uniquely from
experiment in the usual collinear approximation. To any GPD a so-called shadow GPD can be added without changing
the convolutions.

\begin{table*}[t]
   \caption{Moments of the transversity GPD at $t=0$ defined by $K^a_{Tn0}=\int_0^1 dx x^{(n-1)} K_T^a(x,0,0)$
     and comparison with lattice QCD results \ci{LQCD,LQCD2} at the scale $\mu_0$. }  
\renewcommand{\arraystretch}{1.2} 
\begin{center}
  \begin{tabular}{| c  |   c |  c  | c ||  c | c |  c | }
    \hline
                 & Tab. \ref{tab:GPD-parameters} & \req{eq:new-HT} & \ci{LQCD}  & & Tab. \ref{tab:GPD-parameters} & \ci{LQCD2} \\[0.2em]
    \hline
    $H^u_{T10}$   & 0.83 & 0.90 & 0.857(13)  &  $\bar{E}^u_{T10}$ & 3.35 & 2.93(13) \\[0.2em]
    $H^u_{T20}$   & 0.17 & 0.19 & 0.268(6)   &  $\bar{E}^u_{T20}$ & 0.60 & 0.420(31) \\[0.2em]
    $H_{T10}^d$   & -0.05 & -0.06 & -0.212(5)  &  $\bar{E}^d_{T10}$ & 2.03   & 1.90(9)   \\[0.2em]
    $H^d_{T20}$   & -0.007 & -0.007 & -0.052(2)  &  $\bar{E}^d_{T20}$ & 0.32  & 0.260(23) \\[0.2em]
    \hline 
\end{tabular}
\end{center}
\label{tab:moments}
\renewcommand{\arraystretch}{1.0}   
\end{table*}

 \subsection{ The 3-body twist-3 \da}
 For the twist-3 DA of the $q\bar{q}g$ pion's Fock component we use an ansatz advocated for in
 \ci{braun-filyanov}
 \ba
  \phiThreeP(\tau_a,\tau_b,\tau_g) &=& 360 \tau_a\tau_b\tau_g^2 \Big[ 1 + \omega_{1,0}\frac12 (7\tau_g-3)  \nn\\
              && + \omega_{2,0} (2 - 4\tau_a\tau_b - 8\tau_g + 8\tau_g^2)  \nn\\
              && + \omega_{1,1} (3\tau_a\tau_b - 2\tau_g + 3\tau_g^2) \Big]\,.
\label{eq:3-body-DA}
\ea
The DA is normalized as
\be
\int_0^1 d\tau\,\int_0^{\taub} d\tau_g \phiThreeP(\tau,\taub-\tau_g,\tau_g)\=1
\, ,
  \ee
  which goes together with the normalization constant $f_{3\pi}$. This constant as well as the conformal-expansion
  coefficients, $\omega_{i,j}$, depend on the factorization scale, $\mu_F$. The corresponding anomalous dimensions
  can be found in \ci{braun-filyanov} or in \ci{KPK21}. 
Note that $\omega_{2,0}$ and $\omega_{1,1}$ mix under  evolution.

\begin{figure}[t]
\begin{center}
  \includegraphics[width=0.35\tw]{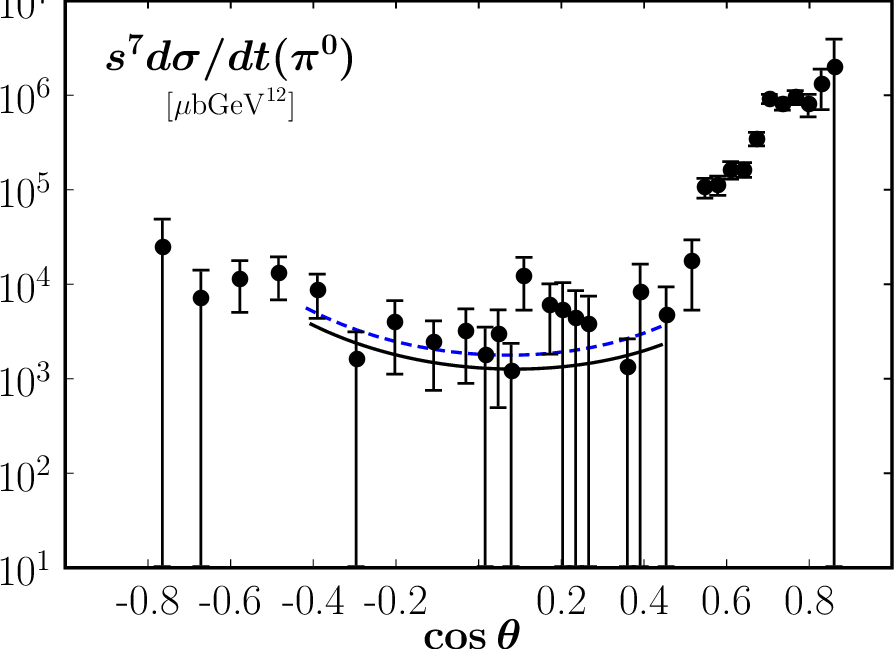}
\end{center}
  \caption{The cross section of $\pi^0$ photoproduction versus the cosine 
of the scattering angle in the center-of-mass system
  at $s=11.06\,\gev^2$. The solid (dashed) line represents the results obtained with
  the 3-body \da{} \req{eq:3-body-da2} (\req{eq:3-body-da1}) and the parameter $A$ controlling the large-$x$ behavior
  of the GPDs 
  $H_T$ and $\bar{E}_T$ 
(see footnote \ref{footnote4})
is chosen to be $0.1 (0.5)\,\gev^{-2}$. The data are taken from \ci{kunkel17}.}
    \label{fig:wa-data}
\end{figure}

  In \ci{KPK18} the normalization constant, $f_{3\pi}$, 
and the coefficient $\omega_{1,0}$ have been taken from a QCD
  sum rule analysis \ci{ball} whereas $\omega_{1,1}$ is assumed to be zero at the initial scale $\mu_0=2\,\gev$,
 and $\omega_{2,0}$
  is fixed by a fit to the wide-angle $\pi^0$ photoproduction data \ci{kunkel17}
(i.e. photoproduction at large Mandelstam variables $s$, $-t$ and $-u$):
  \ba
  f_{3\pi}(\mu_0)&=& 0.004\,\gev^2\,, \qquad \omega_{1,0}(\mu_0)\=-2.55\,,  \nn\\
  \omega_{2,0}(\mu_0)&=& 8.0\,, \hspace*{0.18\tw} \omega_{1,1}(\mu_0)\=0\,.
  \label{eq:3-body-da1}
  \ea
  For reasons which will become clear below we will also
  use a second set of expansion coefficients, namely
  \be
  \omega_{1,0}(\mu_0)\=2.5\,, \qquad \omega_{2,0}(\mu_0)\=6.0\,, 
   \qquad \omega_{1,1}(\mu_0)\=0\,.
  \label{eq:3-body-da2} 
  \ee
  The constant $f_{3\pi}$ remains unaltered. 
The expansion coefficients \req{eq:3-body-da2} 
also provide a reasonable fit
  to the $\pi^0$ photoproduction data,  see Fig.\ \ref{fig:wa-data}. Since the present data on wide-angle
  photoproduction do not fix more than one expansion parameter other sets of expansion coefficients are possible.

In our work within the MPA framework, we also consider quark transverse momenta in the meson. 
Instead of DAs hadron wave functions are required in this case.
  Analogously to the proton wave function \ci{sotiropoulos,bolz} we are writing
  the light-cone wave function of the pion's $q\bar{q}g$ Fock component as
  \be
  \PsiThreeP\= f_{3\pi} \phiThreeP(\tau_1,\tau_2,\tau_g)\, \OmegaThreeP(\vk{}_1,\vk{}_2,\vk{}_g)\,.
\label{eq:wf3}
  \ee
  In the zero-binding limit which is characteristic of the parton picture one has
  \be
  \tau_1+\tau_2+\tau_g\=1\,, \qquad \vk{}_1+\vk{}_2+\vk{}_g\=0\,.
  \label{eq:zero-binding}
  \ee
  The $\vk$-dependence of the wave function \req{eq:wf3} is assumed to be a simple Gaussian with a transverse
  size parameter $\aThreeP$:
  \be
  \OmegaThreeP\= (16\pi^2)^2 \frac{\aThreeP^4}{\tau_1\tau_2\tau_g}\,
                      \exp{[-\aThreeP^2(k^2_{\perp 1}/\tau_1+k^2_{\perp 2}/\tau_2+k^2_{\perp g}/\tau_g)]}\,.
  \label{eq:omega3}
  \ee
  It can readily be seen that
  \be
  \int \frac{d^2\vk{}_1 d^2\vk{}_2 d^2\vk{}_g}{(16\pi^3)^2} \delta^{(2)}(\vk{}_1+\vk{}_2+\vk{}_g)\,\OmegaThreeP\=1\,.
  \ee
  The Fourier transform to the impact parameter plane with respect to the transverse momenta $\vec{k}_{\perp 1}$ and
  $\vec{k}_{\perp 2}$, defined by
  \be
  \hat{f}({\bf b})\=\frac1{(2\pi)^4} \int d^2\vk{}_1 d^2\vk{}_2
                                         \exp{[-i{\bf b}_1\cdot\vk{}_1-i{\bf b}_2\cdot\vk{}_2]} f(\vk)\,,
  \ee
  of the wave function reads
  \be
  \hat{\Psi}_{3\pi}\= f_{3\pi} \phiThreeP \hOmegaThreeP\,,
  \ee
  where
  \be
  \hOmegaThreeP(\vec{b}_1,\vec{b}_2) \= (4\pi)^2\exp{\left\{-\frac1{4\aThreeP^2}\Big[\tau_1\tau_g b_1^2 + \tau_2\tau_g b_2^2
                                     + \tau_1\tau_2 b_g^2\Big]\right\}}\,.
  \label{eq:omega-12}
  \ee
  The transverse separation ${\bf b}_g$ is ${\bf b}_1-{\bf b}_2$. Now one sees that the variable ${\bf b}_1$
  (${\bf b}_2$) is the transverse separation between the quark (antiquark) and the gluon. 
  The Fourier transform with respect to $\vec{b}_1$ and $\vec{b}_g$ is obtained from \req{eq:omega-12}
  by the simultaneous replacement
  \be
  \tau_2\leftrightarrow \tau_g  \qquad \vec{b}_2 \leftrightarrow \vec{b}_g
\, ,
  \ee
  which results in
  \be
  \hOmegaThreeP(\vec{b}_1,\vec{b}_g) \= (4\pi)^2\exp{\left\{-\frac1{4\aThreeP^2}\Big[\tau_1\tau_2 b_1^2 + \tau_2\tau_g b_g^2
                                     + \tau_1\tau_g (\vec{b}_g-\vec{b}_1)^2\Big]\right\}}\,.
  \label{eq:omega-13}
  \ee
  The Fourier transform with respect to $\vec{b}_2$ and $\vec{b}_g$ is obtained analogously.

  In our numerical studies we choose the transverse size parameter $\aThreeP=0.3\,\gev^{-2}$. 
This leads to about the same root-mean-square (rms) 
value of $b_1(=b_2)$, as for the 2-body twist-3 Fock component. 
  
 \subsection{The 2-body twist-3 \da}
 The 2-body twist-3 DA of the pion, $\phiPp$, is uniquely fixed by the 3-body twist-3 DA via the
 equation of motion which, in light-cone gauge, is a first-order linear differential equation \ci{KPK18}.
 Thus, the DA \req{eq:3-body-DA} leads a truncated Gegenbauer expansion of $\phiPp$:
 \be
 \phiPp(\tau)\=1 + \frac17 \frac{f_{3\pi}}{f_\pi\mu_\pi}\,\omega
                 \Big( 10 C_2^{1/2}(2\tau-1) -3 C_4^{1/2}(2\tau-1) \Big)\,,
\label{eq:2-body-tw3-da}
\ee
where
\be
\omega \= 7\omega_{1,0} -2\omega_{2,0} -\omega_{1,1}\,,
\label{eq:omega}
\ee
and $f_\pi$ is the usual pion decay constant for which we take $f_\pi=0.132\,\gev$. 
As usual this DA respects the constraint
\be
\int_0^1 d\tau \phiPp(\tau)\=1\,,
\ee 
and can be written in a more compact form
\be
\phiPp(\tau)\= 1 + \frac{f_{3\pi}}{f_\pi\mu_\pi}\,\omega (1-30\tau^2\taub^2)\,.
\label{eq:2-body-DA}
\ee
In the WW approximation $\phiThreeP$ is zero and $\phiPp$ reduces to $\phiPp^{WW}=1$. There is a second
2-body twist-3 DA, $\phi_{\pi \sigma}$, which is also fixed by the 3-body \da{} via the equation of motion. We
do not quote it here because it does not contribute to DVMP as has been shown in \ci{GK5}.

\begin{figure}[t]
\begin{center}
  \includegraphics[width=0.35\tw]{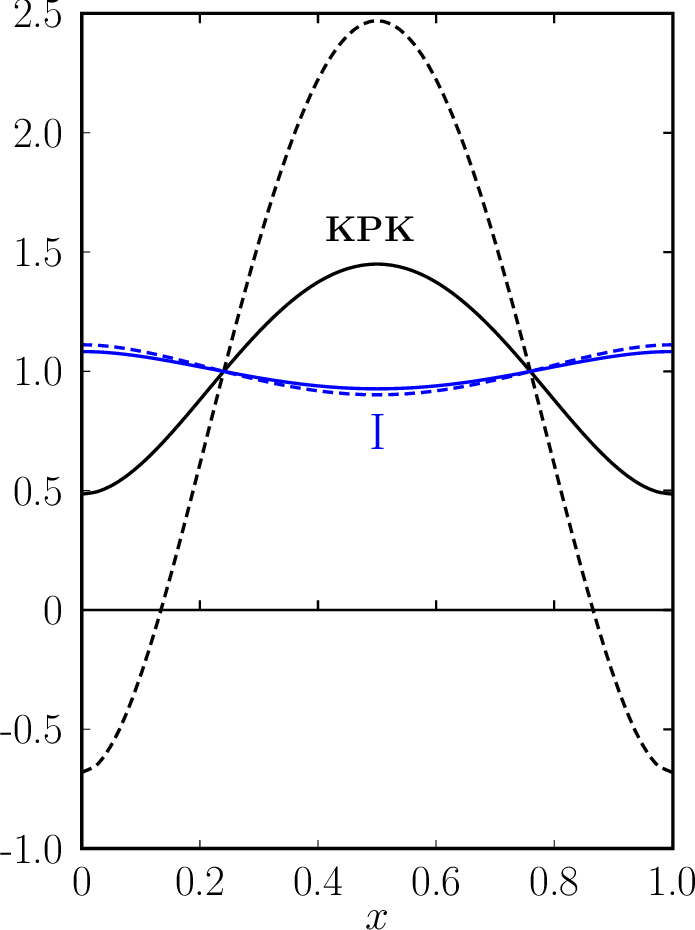}
\end{center}
  \caption{The DA $\phiPp$ vs. $x$. The solid 
line denoted KPK (I) represents the DA generated
    from the expansion coefficients \req{eq:3-body-da1} 
(\req{eq:3-body-da2}) at the scale $\mu_F=2.0\,\gev$.
    The dashed lines are the DAs at the scale $\mu_F=0.7\,\gev$.}
    \label{fig:DA}
\end{figure}

It is inspiring to examine the evolution behavior of $\phiPp$:
The mass parameter evolves with the scale as
\be
\mu_\pi(\mu_F)\=L^{-4/\beta_0}\, \mu_\pi(\mu_0)\,,
\ee
where $\beta_0=(11N_C-2n_f)/3$ and
\be
L\=\frac{\ln(\mu_0^2/\LQCD^2)}{\ln(\mu_F^2/\LQCD^2)}\,.
\ee
It follows that $\mu_\pi$ is small at small scales and becomes large for $\mu_F\to \infty$. 
This untypical evolution behavior is caused by the current quark masses in the denominator 
of $\mu_\pi$, see \req{eq:mupi}. 
On the other hand,
\ba
f_{3\pi}(\mu_F)&=&L^{(16/3C_F-1)/\beta_0}\, f_{3\pi}(\mu_0)\,,\nn\\
\omega_{1,0}(\mu_F) &=& L^{(-25/6C_F+11/3C_A)/\beta_0}\,\omega_{1,0}(\mu_0)\,,
\ea
while a more complicated scale dependence of $\omega_{2,0}$ and $\omega_{1,1}$ 
is a consequence of their mixing under evolution, see \ci{KPK21,braun-filyanov}. 
Their scale dependence is similarly strong as that of $\omega_{1,0}$.
Now we understand the evolution behavior of the second term of $\phiPp$: it is large at small scales but tends
to zero for $\mu_F\to\infty$. In Fig.\ \ref{fig:DA} we display $\phiPp$ generated by the two 3-body DAs
\req{eq:3-body-da1} and \req{eq:3-body-da2}. 
The combination $\omega$ strongly differs in magnitude and sign for the two cases
\be
\omega(\mu_0) \= -33.85
\, ,
\label{eq:omega1}
\ee
for \req{eq:3-body-da1} and
\be
\omega(\mu_0) \= 5.5
\, ,
\label{eq:omega2}
\ee
for  \req{eq:3-body-da2}. These different values of $\omega$ lead to a drastically different behavior of $\phiPp$ at
low scales as Fig.\ \ref{fig:DA} reveals. The first value of $\omega$ leads to a $\phiPp$ with
pronounced maxima and minima whereas the DA, evaluated from \req{eq:omega2}, remains close to unity, i.e.\
close to $\phiPp^{WW}$. Only for very low scales close to $\LQCD$, this DA differs substantially from its
WW approximation.

In the following we will also need a light-cone wave function for 2-body twist-3 Fock component of the pion for which
we will use \ci{GK5,GK6}
\be
\PsiPp\= \frac{16\pi^{3/2}}{\sqrt{2N_C}}\,f_\pi \aPp^3 k_\perp \phiPp(\tau) \exp{[-\aPp^2k_\perp^2]}
\,.
\label{eq:2-body-tw3-WF}
\ee
For the transverse size parameter, the value $\aPp=1.8\,\gev^{-1}$ has been used in \ci{GK5,GK6} and will be applied
by us as well. 
This value of $\aPp$ corresponds to a rms value
of $\sqrt{\langle b^2\rangle}=0.5$\,fm. 

It is easy to show that
for integer $n$
\be
\int \frac{d^2\vk}{16\pi^3}\,k_\perp^{2n}\PsiPp \= \frac{f_\pi}{2\sqrt{2N_C}}\,\phiPp\, \aPp^{-2n}
                                                  \frac{2}{\sqrt{\pi}} \Gamma(n+3/2)
\,.
  \label{eq:int-2}
  \ee

\section{The modified perturbative approach}
\label{sec:MPA}
As we already mentioned, the 2-body twist-3 subprocess amplitude \req{eq:2-body-tw3-sub} possesses an end-point
singularity. Following \ci{GK5} where the WW approximation of this amplitude has been applied, we are going
to calculate the subprocess amplitudes within the MPA in which
transverse momenta of the partons entering the pion are taken into account.
The emission and reabsorption of quarks by the nucleons is still treated collinearly to the nucleon momenta.
This scenario is justified to some extent by the fact that the GPDs describe the full proton, and their
$\vk$-dependence therefore reflects the nucleon charge radius ($\langle k_\perp^2\rangle^{1/2}\simeq 200\,\mev$),
while the pion is mainly generated through its compact $q\bar{q}$ Fock component with a r.m.s. $k_\perp$ of about
$500\,\mev$. The \da s of the collinear approximation are to be replaced by light-cone wave function in this
scenario. The parton transverse momenta are accompanied by gluon radiation. In \ci{botts89} the gluon radiation
has been calculated in form of a Sudakov factor $\exp{[-S]}$ to next-to-leading log approximation using resummation
techniques and having recourse to the renormalization group. Since the resummation of the logarithms involved in the
Sudakov factor can only be efficiently performed in the impact parameter space \ci{botts89} we have to work
in that space. The Sudakov factor is zero for $b\geq 1/\LQCD$. This cut-off generates a series of power suppressed
terms which come from the region of soft quark momenta. The interplay of the quark transverse momenta and the
Sudakov factor regularizes the above mentioned end-point singularity. For more details see \ci{GK3}.

\subsection{The 2-body twist-3 case}
We assume that the quark and antiquark momenta of the pion's constituents are
\be
\tau q' + K\,,        \qquad \taub q' - K\,,
\label{eq:parton-momenta}
\ee
where $q'$ is the momentum of the pion and
\be
q'\cdot K\=0\,,   \qquad K^2\=-k_\perp^2\,.
\label{eq:transverse}
\ee
The leading-order (LO) perturbative calculation 
reveals that the $k_\perp$-dependence appears only in the gluon propagator
and the double poles in \req{eq:2-body-tw3-sub} become
\be
\frac1{(\sh +i\eps)(\taub\sh-k_\perp^2+i\eps)}\,,   \qquad \frac1{(\uh+i\eps)(\tau\uh-k_\perp^2+i\eps)}\,.
\ee
We decompose the product of propagators into a sum of two single propagators
\ba
\frac1{(\sh +i\eps)(\taub\sh-k_\perp^2+i\eps)}&=& \frac1{k_\perp^2}\Big[-\frac1{\sh+i\eps}
                                                          + \frac{\taub}{\taub\sh-k_\perp^2+i\eps}\Big] 
\, ,
\nn\\
\frac1{(\uh+i\eps)(\tau\uh-k_\perp^2+i\eps)}&=&   \frac1{k_\perp^2}\Big[-\frac1{\uh+i\eps}
  + \frac{\tau}{\tau\uh-k_\perp^2+i\eps}\Big]\,.
\label{eq:partial}
\ea
Using the $\tau \leftrightarrow \taub$ symmetry of $\phiPp$ and the relation \req{eq:int-2} as well as
the above decomposition, we can write the subprocess amplitude \req{eq:2-body-tw3-sub} as
\ba
   {\cal H}^{\pi^i,q\bar{q}}_{0-\lambda,\mu\lambda}&=& 
-2\sqrt{2}\pi (2\lambda+\mu) {\cal C}_{\pi^i}^{(ab)} f_\pi\mu_\pi
   \aPp^2\als(\mu_R) \frac{C_F}{N_C} \frac{Q^2}{\xi} 
\nn\\  &&\times 
\Big[\frac{e_a}{\sh+i\eps} + \frac{e_b}{\uh+i\eps}\Big] \nn\\
   &&+\: 4\pi (2\lambda+\mu)  {\cal C}_\pi^{(ab)} \mu_\pi \frac{C_F}{\sqrt{N_C}} \frac{Q^2}{\xi}\,
                \int d\tau \int \frac{d^2\vk}{16\pi^3} k_\perp^{-2} \PsiPp \als(\mu_R) \nn\\
          &&\times  \Big[\frac{\taub e_a}{\taub \sh  - k_\perp^2 + i\eps} 
                  +      \frac{\tau e_b}{\tau \uh  - k_\perp^2 + i\eps} \Big]\,.
 \label{eq:inter-P}               
 \ea
 The next step is to transform the subprocess amplitude to the impact parameter plane.
 Since the wave function appears to be divided by $k_\perp^2$ it is convenient to transform the product
 $k_\perp^{-2} \PsiPp$. Using the wave function \req{eq:2-body-tw3-WF} this Fourier transform is
 \be
    [k_\perp^{-2} \PsiPp]_{\rm FT}\= 4\pi \frac{f_\pi \aPp^2}{\sqrt{2N_C}} \phiPp(\tau)
                     I_0(b^2/8a_p^2) e^{-b^2/(8\aPp^2)}\,.
\ee
Here, $I_0$ is the Bessel function of order zero. 
Replacing $k_\perp^{-2} \PsiPp$ in \req{eq:inter-P} 
by its Fourier transform
\be
k_\perp^{-2} \PsiPp\= \int d^2{\bf b}\, e^{-i{\bf b}\cdot \vk}\,[k_\perp^{-2} \PsiPp]_{\rm FT}
\, ,
\ee
it remains to perform the Fourier transform of the propagators which can easily be done using
\be
\int\frac{d^2\vk}{(2\pi)^2}\,\frac{e^{-i\vk\cdot {\bf b}}}{\vk^2-a-i\eps} \= \frac{i}{4}\,H_0^{(1)}(\sqrt{a}b)\,\Theta(a)
+\frac1{2\pi}\,K_0(\sqrt{-a}b)\,\Theta(-a)\,,
\label{eq:master}
\ee
where $H_0^{(1)}$ and $K_0$ denote Hankel and Bessel functions of the second kind, respectively.
Thus, we finally arrive at
\ba
   {\cal H}^{\pi^i,q\bar{q}}_{0-\lambda,\mu\lambda}&=& 
-2\sqrt{2}\pi (2\lambda+\mu) {\cal C}_{\pi^i}^{(ab)} f_\pi\mu_\pi
   \aPp^2 \frac{C_F}{N_C} \frac{Q^2}{\xi}
\nn\\ &&\times \left\{\als(\mu_R) \Big[\frac{e_a}{\sh+i\eps} + \frac{e_b}{\uh+i\eps}\Big] \right.\nn\\
   &&+\left.  \int d\tau \phiPp (\tau) \int bdb I_0(b^2/(8\aPp^2)) e^{-b^2/(8\aPp^2)}
                     \als(\mu_R) e^{-S} \right. \nn\\
  &&\times \left. \Big[ \taub e_a \Big(i\frac{\pi}{2} H_0^{(1)}(\sqrt{\taub \sh}\, b)\, \Theta(\sh)\,
                      + \,K_0(\sqrt{-\taub \sh}\, b)\, \Theta(-\sh) \Big) \right.\nn\\
                       && \left. \quad + \: \tau e_b K_0(\sqrt{-\tau \uh}\, b)\, \Theta(-\uh) \Big] \right\}\,.
  \label{eq:2-body-tw3-sub-amp}                   
  \ea
  This subprocess amplitude is to be convoluted with a transversity GPD in accordance with \req{eq:hel-amp}.
  In the spirit of the MPA we have added the Sudakov factor $\exp{[-S(\tau, {\bf b}, Q^2)]}$ under the integral.
  The quark-antiquark separation, ${\bf b}$, in the impact parameter space acts as an infrared cut-off.
  Radiative gluons with wave lengths larger than the infrared cut-off are part of the pion wave function.  
  Those gluons with wave lengths between the infrared cut-off and a lower limit (related to the hard scale $Q^2$)
  yield suppression while harder  ones are part of the perturbative subprocess amplitude. In this situation the
  factorization scale is naturally given by $\mu_F=1/b$. 
With regard to the scale dependence of the DA $\phiPp$
  we stop the evolution at $0.5\,\gev$, i.e.\ $\mu_F={\rm max}(1/b, 0.5\,\gev)$. 
Not all logarithmic singularities 
  arising from the evolution of the DA $\phiPp$ are canceled by the Sudakov factor as was the case for
  the WW approximation used in \ci{GK5,GK6}.
  The renormalization scale is taken to be the largest scale appearing in the subprocess,
  $\mu_R={\rm max}(\tau Q, \taub Q, 1/b)$. 
  
\subsection{The 3-body case}
We have now to deal with three parton transverse momenta defined analogously to Eqs.\ \req{eq:parton-momenta} and
\req{eq:transverse} and which satisfy the condition \req{eq:zero-binding}.
From the LO perturbative calculation of the subprocess amplitudes \req{eq:3-body-tw3-CF} and
\req{eq:3-body-tw3-CG} we learn that always
two different parton transverse momenta appear in the propagators in contrast to the 2-body case
\footnote{As explained in \ci{GK6}, 
in the spirit of MPA we only retain $k_\perp$ in the denominators of the propagators
where both momentum fractions $x$ and $\tau$ appear.}. 
We expect therefore a strong suppression of the 3-body contributions.
Introducing the 3-body wave function \req{eq:wf3} instead of the distribution amplitude as we did analogously
for the 2-body twist-3 case, the subprocess amplitude ${\cal H}^{\pi^i,q\bar{q}g, C_F}_{0-\lambda,\mu\lambda}$ in Eq.\
\req{eq:3-body-tw3-CF} reads
\ba
          {\cal H}^{\pi^i,q\bar{q}g,C_F}_{0-\lambda,\mu\lambda} &=& -(2\lambda+\mu) \sqrt{2}\pi {\cal C}^{(ab)}_P\frac{C_F}{N_C}
          \frac{Q^2}{\xi} \int_0^1 \frac{d\tau_1}{\taub_1}\int_0^{\taub_1} d\tau_g
         \int \frac{d^2k_{\perp 1}d^2k_{\perp 2}}{(16\pi^3)^2}\PsiThreeP \nn\\
         &&\times \: \als(\mu_R) \left\{e_a\frac{1}{(\taub_1\sh-k^2_{\perp 1}+i\eps)(\tau_2\sh
                 -k^2_{\perp 2}+i\eps)} \right. \nn\\
         &&\left.\hspace*{0.13\tw}   + e_b\frac{1}{(\taub_1\uh-k^2_{\perp 1}+i\eps)(\tau_2\uh-
                k^2_{\perp 2}+i\eps)} \right\}
\, .
       \label{eq:CF-ampl-kperp}
       \ea
Transforming to the impact parameter space and using \req{eq:master} we arrive at
 \be
      {\cal H}^{\pi^i,q\bar{q}g,C_F}_{0-\lambda,\mu\lambda}\=- (2\lambda +\mu) \sqrt{2}\pi {\cal C}_P^{(ab)}\frac{C_F}{N_C}
      \frac{Q^2}{\xi} f_{3\pi} \int_0^1 \frac{d\tau_1}{\taub_1}\int_0^{\taub_1} d\tau_g \phiThreeP
      \,{\cal H}_1
\, ,
 \label{eq:sub-CF} 
      \ee
 where
 \ba     
             {\cal H}_1 &=&  \int b_1db_1 b_2db_2 \exp{[-\frac1{4\aThreeP^2}(\tau_1\taub_1b_1^2+\tau_2\taub_2b_2^2)]}\,
                        \,\als(\mu_R)e^{-S(b_1,b_2)}    \nn\\
      &&\times \: I_0(\tau_1\tau_2b_1b_2/(2\aThreeP^2))\,\left\{e_a \left(-\frac{\pi^2}{4}
                            H_0^{(1)}(\sqrt{\taub_1\sh}\,b_1)\, 
                            H_0^{(1)}(\sqrt{\tau_2\sh}\,b_2)\,\Theta(\sh) \right.\right. \nn\\
     &&+ \left.\left.   K_0(\sqrt{-\taub_1\sh}\,b_1)\,K_0(\sqrt{-\tau_2\sh}\,b_2)\,\Theta(-\sh)
                               \right)\right.\nn\\
      &&+\left. \; e_b K_0(\sqrt{-\taub_1\uh}\,b_1)\,K_0(\sqrt{-\tau_2\uh}\,b_2)\, \Theta(-\uh) \right\}\,.        
 \label{eq:H1}
 \ea
 The angle integrations implied in $d^2b_i$ have already been carried out and the Sudakov factor is introduced.
 The momentum fraction $\tau_2$ is
 \be
 \tau_2\= 1-\tau_1-\tau_g\,.
 \ee

 The subprocess amplitude ${\cal H}^{\pi^i,q\bar{q}g, C_G}_{0-\lambda,\mu\lambda}$ \req{eq:3-body-tw3-CG} is treated
 analogously although it is much more complicated because any pair of parton transverse momenta $\vec{k}_i$,
 $\vec{k}_j$ occur in the propagators:
 \ba
    {\cal H}^{\pi^i,q\bar{q}g,C_G}_{0-\lambda,\mu\lambda}&=& (2\lambda + \mu)\,\sqrt{2}\pi {\cal C}_P^{(ab)}\frac{C_G}{N_C}
                               \frac{Q^2}{\xi} \,\int_0^1\frac{d\tau_1}{\taub_1}\,\int_0^{\taub} d\tau_g \nn\\
                 &&\times \int \frac{d^2k_{\perp 1}d^2k_{\perp 2}d^2k_{\perp g}}{(16\pi^3)^2}
                               \delta(\vec{k}_{\perp 1}+\vec{k}_{\perp 2}+\vec{k}_{\perp g}) \PsiThreeP\als(\mu_R) \nn\\
                  &&\times \Big[\frac{e_a}{(\taub_1\sh-k^2_{\perp 1}+i\eps)(\tau_2\sh-k^2_{\perp 2}+i\eps)} \nn\\
                  && \:+ \frac{e_b}{(\taub_1\uh-k^2_{\perp 1}+i\eps)(\tau_2\uh-k^2_{\perp 2}+i\eps)} \nn\\
              &&\: + \frac{e_a}{(\taub_1\sh-k^2_{\perp 1}+i\eps)(\tau_g\sh-k^2_{\perp g}+i\eps)}   \nn\\                      
              && \: + \frac{e_b}{(\taub_1\uh-k^2_{\perp 1}+i\eps)(\tau_g\uh-k^2_{\perp g}+i\eps)}  \nn\\
              && \: + \frac{e_a}{(\tau_2\sh-k^2_{\perp 2}+i\eps)(\tau_g\uh-k^2_{\perp g}+i\eps)}  \nn\\         
              && \: +   \frac{e_b}{(\tau_2\uh-k^2_{\perp 2}+i\eps)(\tau_g\sh-k^2_{\perp g}+i\eps)}\Big] 
\, .
    \ea
In the impact parameter space we have
 \ba
    {\cal H}^{\pi^i,q\bar{q}g,C_G}_{0-\lambda,\mu\lambda}&=& \sqrt{2}\pi (2\lambda + \mu)\,{\cal C}_P^{(ab)}
    \frac{C_G}{N_C} \frac{Q^2}{\xi} f_{3\pi}  \int_0^1 \frac{d\tau_1}{\taub_1}\,\int_0^{\taub_1} d\tau_g
                             \phiThreeP \nn\\
                  &\times&  \Big[{\cal H}_1 + {\cal H}_2 +{\cal H}_3\Big]
\, ,
\label{eq:sub-CG}
\ea 
where ${\cal H}_1$ is given in \req{eq:H1} while ${\cal H}_2$ is obtained from ${\cal H}_1$ by the replacement
\be
\tau_2 \to \tau_g\,, \qquad b_2\to b_g\,,
\ee
and 
 \ba
    {\cal H}_3 &=& \int b_2db_2b_gdb_g
          \exp{[-\frac1{4\aThreeP^2}(\tau_2\taub_2b_2^2 +\tau_g\taub_gb_g^2)]}
                  I_0(\tau_2\tau_gb_2b_g/(2\aThreeP^2)) \als e^{-S(b_2,b_g)}\nn\\
              &&\times\left\{e_a\Big[i\frac{\pi}{2} H_0^{(1)}(\sqrt{\tau_2\sh}\,b_2) \Theta(\sh)\, 
                  +\, K_0(\sqrt{-\tau_2\sh}\,b_2) \Theta(-\sh) \Big] \right. \nn\\ 
               &&\times \left.  K_0(\sqrt{-\tau_g\uh}\,b_g) \Theta(-\uh) \,
                 + \,e_b K_0(\sqrt{-\tau_2\uh}\,b_2) \Theta(-\uh)\,   \right. \nn\\
                 &&\times\left. \Big[i\frac{\pi}{2} H_0^{(1)}(\sqrt{\tau_g\sh}\,b_g) \Theta(\sh)\,
                 +\,  K_0(\sqrt{-\tau_g\sh}\,b_g) \Theta(-\sh)\Big] \right\}
\, .
    \ea

The Sudakov factor in the above amplitudes 
is that of the $q\bar{q}g$ system which is unknown.
We therefore approximate the Sudakov factor by
\be
e^{-S(b_i,b_j)} \simeq \Theta(b_0-b_i) \Theta(b_0-b_j)
\, ,
\label{eq:qbarqg-sudakov}
\ee
where $b_0=1/\LQCD$.
This way we rather overestimate the 3-body contribution since the Sudakov factor suppresses
the amplitudes already at $b<b_0$. As it turns out from our numerical studies described in  Sec.\
\ref{sec:MPA-results}, the 3-body contribution is much smaller than the 2-body twist-3
one. In fact it is almost negligible. Hence, our approximation suffices. As we already mentioned above the
cut-off of the $b$-integrals generates a series of power suppressed terms which come from the region of soft quark
momenta. We also stress that there are neither end-point singularities in the 3-body contribution nor double poles.

\subsection{Numerical studies}
\label{sec:MPA-results}

In this subsection we are going to compare 
our twist-3 contribution evaluated within the MPA 
to experimental data on deeply virtual electroproduction of pions. 
We will demonstrate that the 
twist-3 contribution
- accompanied by a moderate adjustment of the transversity GPDs - leads to reasonable results for a sample of
kinematical settings $(Q^2, x_B)$ for which experimental data are available from either CLAS \ci{CLAS14},
the Hall A collaboration \ci{hall-A-2020} or COMPASS \ci{COMPASS-2019}. We stress that we restrict ourselves
to $\pi^0$ production because in this case the contributions from transversal photons which are of twist-3 nature,
are dominant. This is to be contrasted with charged pion production, where longitudinal photons play the decisive
role, at least at small $-t'$. This is mainly caused by the large contribution from the pion pole \ci{GK5}.

Let us first discuss the transverse-transverse interference cross section that is defined in terms of
$\gamma^* p\to \pi^0 p$ helicity amplitudes by (for convenience we drop the subscript $\pi^0$ in this subsection)
\be
\frac{d\sigma_{TT}}{dt}\= -\frac{2 {\rm Re}\Big[ {\cal M}^*_{0-,++} {\cal M}_{0-,-+} + {\cal M}^*_{0+,++} {\cal M}_{0+,-+}\Big]}
  {32\pi (W^2-m^2)\sqrt{\Lambda(W^2,-Q^2,m^2)}}
\, ,
\label{eq:sigmaTT}
  \ee
  where $\Lambda$ is the familiar Mandelstam function and $W$ is the energy in the pion-(final-state) proton
  center-of-mass system. The center-of-mass helicity amplitudes are given by the convolutions \req{eq:hel-amp}
  of transversity GPDs and the subprocess amplitudes discussed in the preceding subsection.
  From \req{eq:hel-amp} we see that the first term in \req{eq:sigmaTT} is zero and the
  second one is equal to 
  \be
  \sim |{\cal M}_{0+,++}|^2\,.
  \ee
  We also see from this equation and \req{eq:hel-amp} that only the GPD $\bar{E}_T$ contributes to this
  interference cross section.

\begin{figure}[th]
  \begin{center}
   \includegraphics[width=0.36\tw]{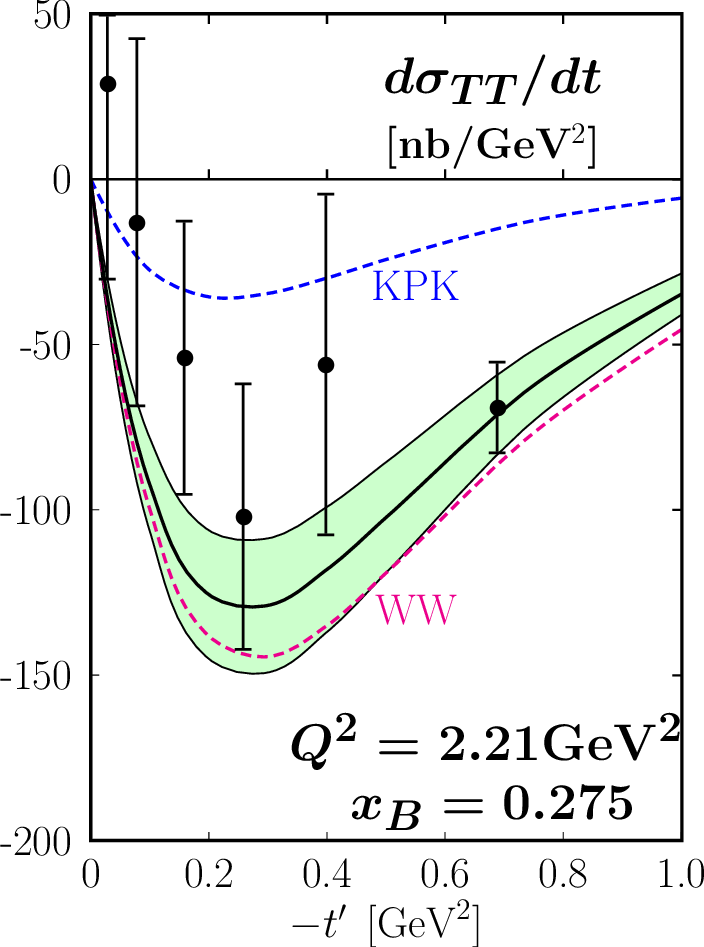} \hspace*{0.05\tw}
\includegraphics[width=0.36\tw]{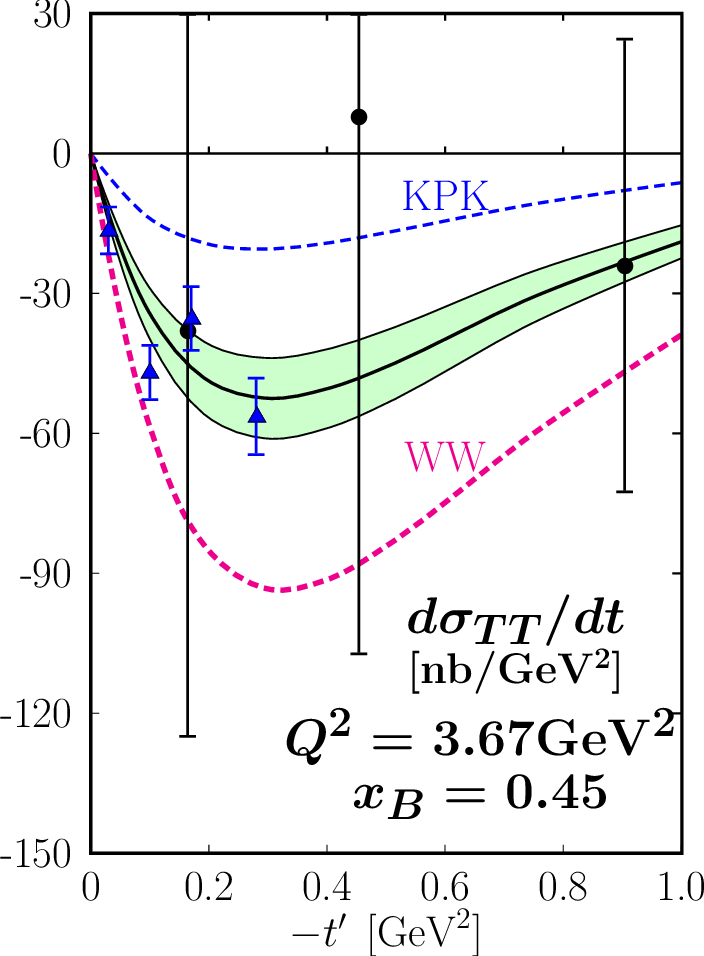}
\includegraphics[width=0.36\tw]{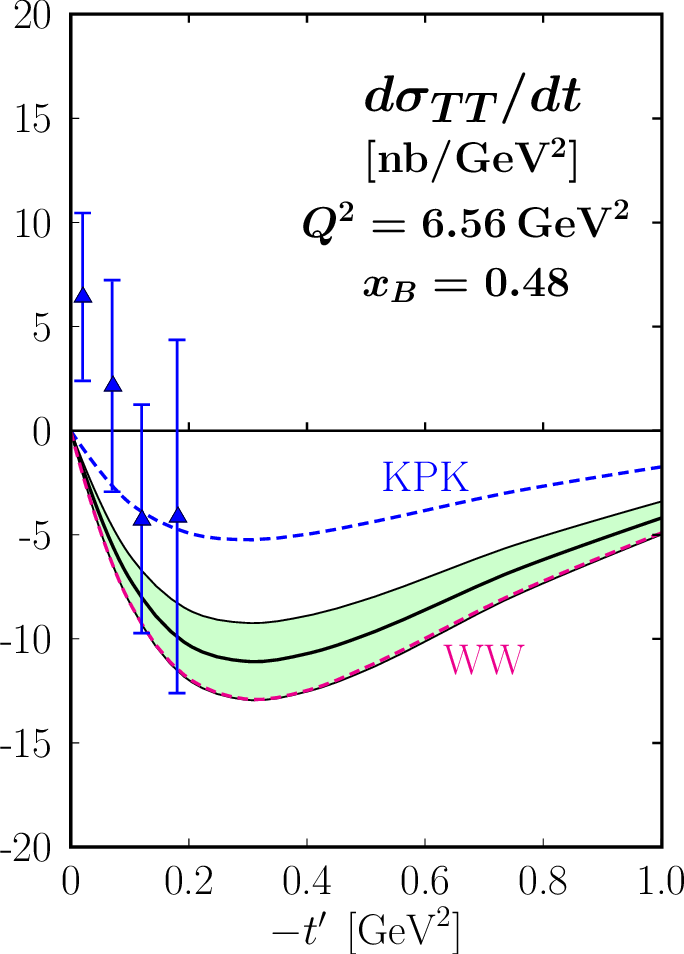}  \hspace*{0.05\tw}
\includegraphics[width=0.36\tw]{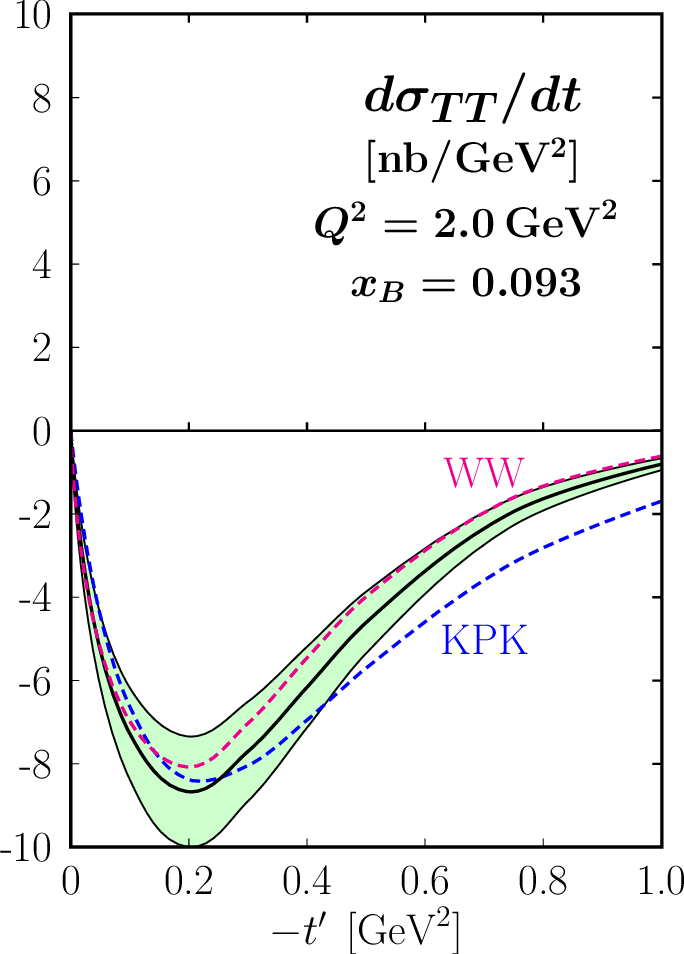}
  \end{center}
  \caption{The transverse-transverse interference cross section versus $t'$ for various
    kinematical settings. The solid lines with error bands 
(evaluated from the uncertainties of the GPDs and $\mu_\pi$)
are the MPA results evaluated from the
    \da{} \req{eq:3-body-da2}, the dashed lines are evaluated from the \da{} \req{eq:3-body-da1} (KPK)
    and from the WW approximation (WW). The latter result is taken from \ci{GK6}. The data are taken from
    \ci{CLAS14} (full circles) and from \ci{hall-A-2020} (triangles). The Hall A data in the upper right plot
  are at the adjacent kinematics $Q^2=3.57\,\gev^2$ and $x_B=0.36$.}
  \label{fig:TT}
\end{figure}

In Fig.\ \ref{fig:TT} we display our results evaluated from the 3-body twist-3 \da{} \req{eq:3-body-da1}
and the corresponding 2-body twist-3 \da{} \req{eq:2-body-DA} (termed KPK in the following), 
and similarly from the DA \req{eq:3-body-da2}. 
For comparison we also show the results obtained with the WW approximation 
\ci{GK6}. The various results are evaluated from the GPDs defined in Tab.\ \ref{tab:GPD-parameters}.
As is evident from Fig.\ \ref{fig:TT} the results on $d\sigma_{TT}$ obtained from the \da{} \req{eq:3-body-da2} are
in very good agreement with experiment. The KPK results, on the other hand, are a bit worse.

\begin{figure}[th]
  \begin{center}
   \includegraphics[width=0.36\tw]{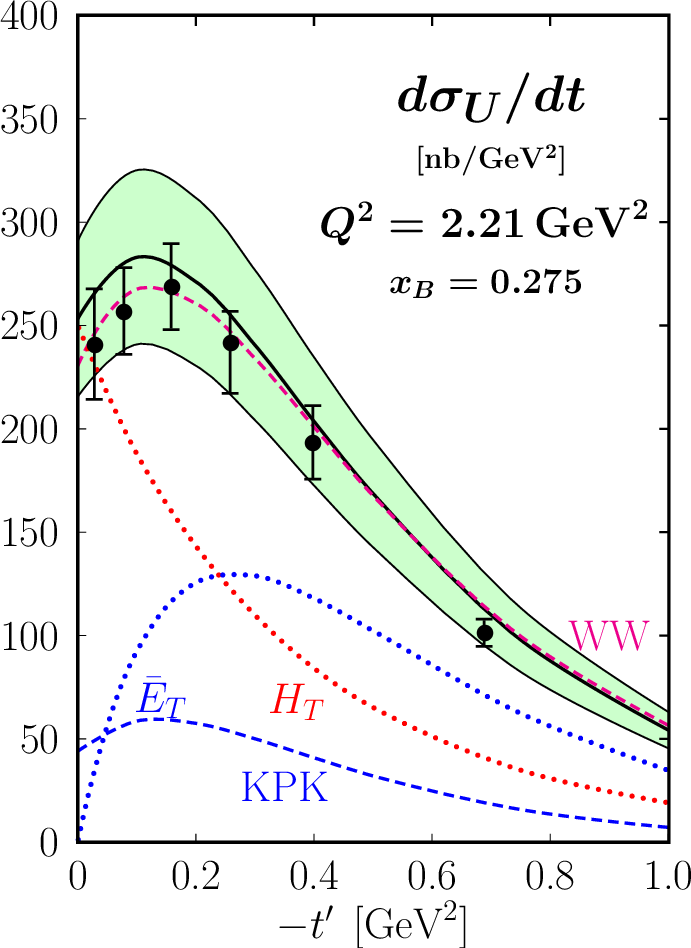} \hspace*{0.05\tw}
\includegraphics[width=0.36\tw]{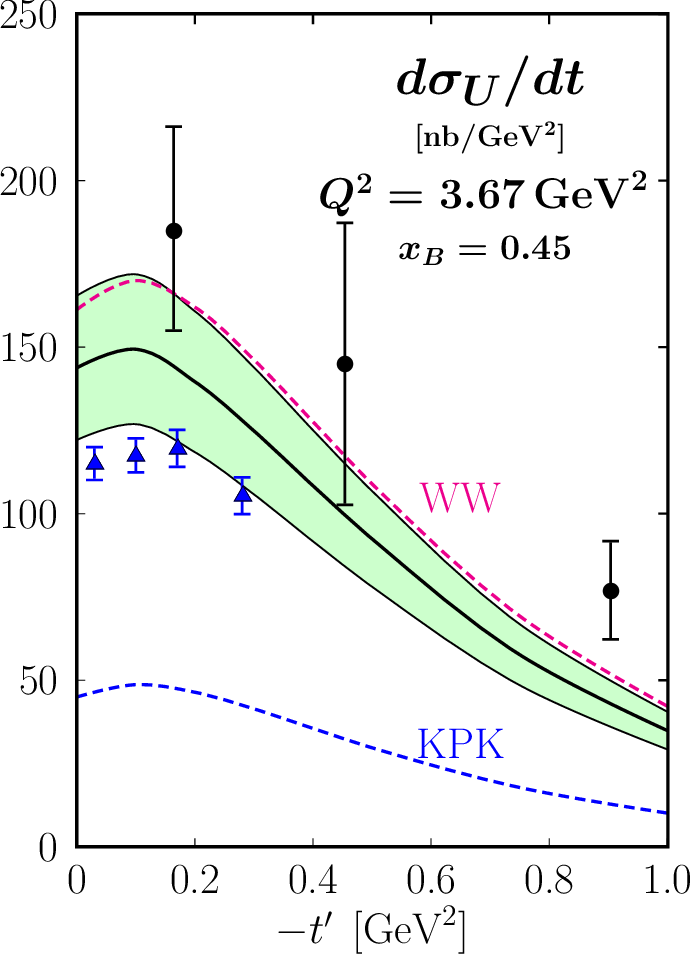}
\includegraphics[width=0.36\tw]{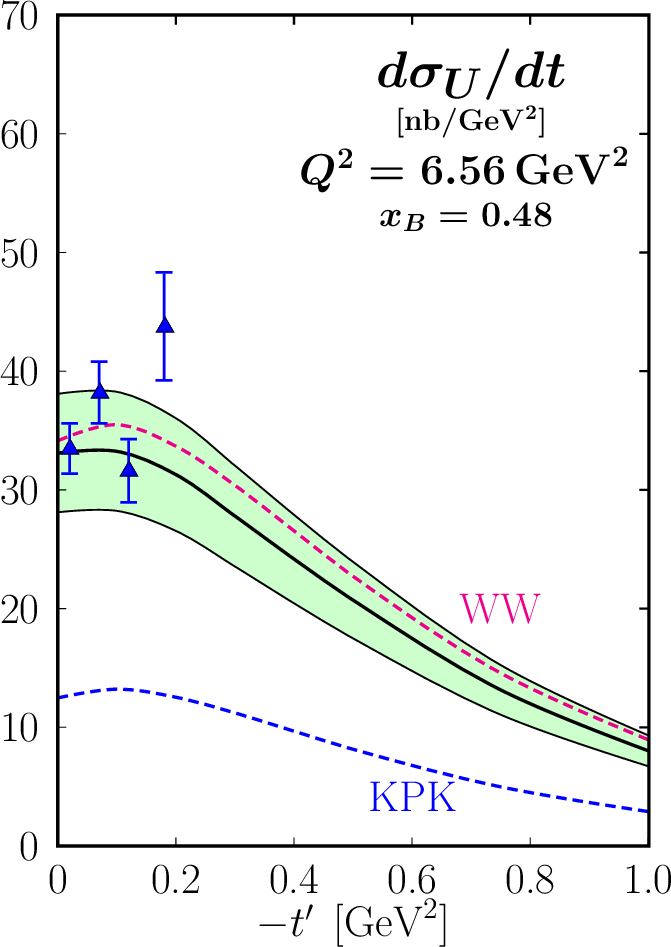}  \hspace*{0.05\tw}
\includegraphics[width=0.36\tw]{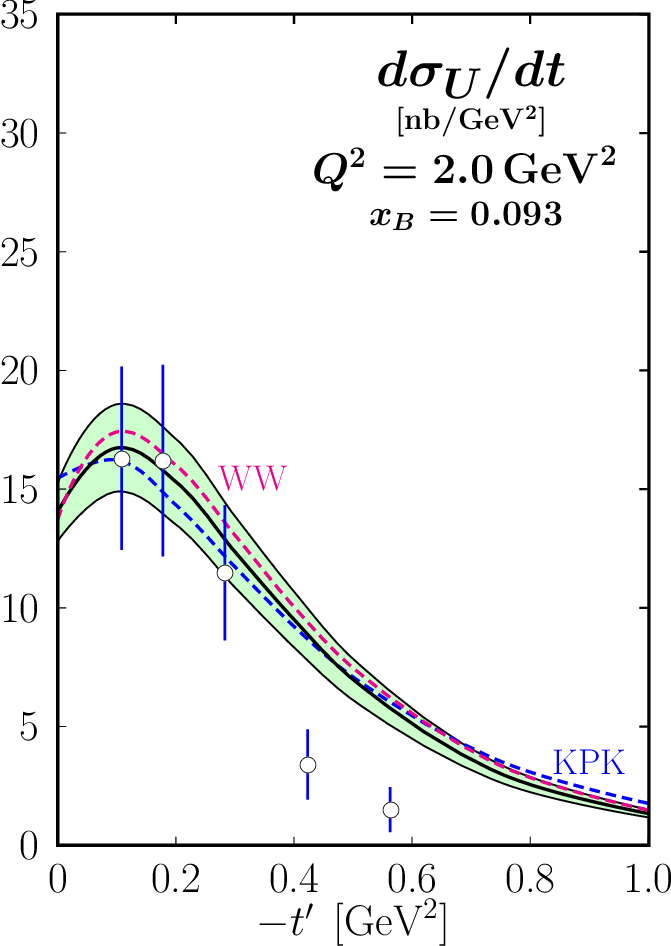}
  \end{center}
  \caption{The unseparated cross section versus $t'$ for various kinematical settings. The separate contributions
    from $H_T$ and $\bar{E}_T$ are shown as dotted lines for $Q^2=2.21\,\gev^2$. The data shown by open circles
    are taken from \ci{COMPASS-2019}. 
For other notations, see Fig.\ \ref{fig:TT}.}
  \label{fig:T}
\end{figure}

In Fig.\ \ref{fig:T} we show the results for the unseparated cross section defined by
\be
\frac{d\sigma_U}{dt} \= \frac{d\sigma_T}{dt} + \eps \frac{d\sigma_L}{dt}
\, ,
\label{eq:sigU}
\ee
where $\eps$ is the ratio of the longitudinal and transversal polarization of the virtual
photon and $d\sigma_L$ is the longitudinal cross section that is fed by the twist-2
subprocess amplitudes and the GPDs $\widetilde{H}$ and $\tilde{E}$, see Tab.\ \ref{tab:GPD-parameters}.
We take the longitudinal cross section from \ci{GK6}. It is small, about $3\%$ of the transverse
cross section at $x_B=0.275$ and even smaller for larger $x_B$. The smallness of the predicted longitudinal
cross section at an $x_B$ of about 0.3-0.4  is in agreement with experiment \ci{defurne,mazouz}.
At the COMPASS kinematics however $d\sigma_L$ is substantially larger. It amounts to about $40\%$ of the
transverse cross section. Mainly responsible for the increase of the ratio $d\sigma_L/d\sigma_T$ with
decreasing Bjorken-$x$ (at fixed $Q^2$) is the GPD parameter $\alpha(0)$ (see Tab.\ \ref{tab:GPD-parameters}) 
which acts like a Regge intercept. 
For $x_B\to 0$ (at fixed $Q^2$) a GPD contributes to a cross section as
\be
d\sigma \sim x_B^{-2(\alpha(0)-1)}
\, .
\ee
While $\alpha(0)$ is about 0.3-0.5 for the GPDs $\widetilde{H}$ and $\tilde{E}$ contributing
to the longitudinal cross section, for the transversity GPDs its value is about -0.2 and -0.1 (see
Tab.\ \ref{tab:GPD-parameters} and Eq.\ \req{eq:HT-parameterization}).

In terms of the amplitudes \req{eq:hel-amp} the transverse cross section reads
\be
\frac{d\sigma_T}{dt}\= \frac{|{\cal M}_{0-,++}|^2 + 2 |{\cal M}_{0+,++}|^2}
                                    {32\pi (W^2-m^2) \sqrt{\Lambda(W^2,-Q^2,m^2)}}\,.
\label{eq:sigmaT}
\ee
\clearpage
\clearpage
Comparing \req{eq:sigmaTT} and \req{eq:sigmaT} one notices the bound
\be
\left|\frac{d\sigma_{TT}}{dt}\right| \leq \frac{d\sigma_T}{dt}
\, ,
\ee
which holds generally not only for the amplitudes \req{eq:hel-amp}. Comparison of the data shown in Figs.\
\ref{fig:TT} and \ref{fig:T}  makes it clear that the transversal cross section, under control of the twist-3
contributions and the GPD $\bar{E}_T$, amounts to a substantial fraction of the unseparated cross section.
Both the GPDs, $H_T$ and $\bar{E}_T$, contribute to the transverse cross section. The $H_T$ contribution
dominates at small $-t'$, $\bar{E}_T$ at larger $-t'$, see Fig.\ \ref{fig:T}.

Inspection of Fig.\ \ref{fig:T} reveals that the KPK results obtained from the
GPDs quoted in Tab.\ \ref{tab:GPD-parameters}, are much smaller than experiment.
Evaluating instead the transverse cross section from the DA \req{eq:3-body-da2} leads to results that
are very close to experiment. Still they can be improved by changing the normalization of $H_T$
\be
N_{H_T}^u\=1.17\,, \hspace*{0.1\tw} N_{H_T}^d\=-0.31\,,
\label{eq:new-HT}
\ee
with a corresponding change of the moments, see Tab.\ \ref{tab:moments}. The  agreement with the lattice QCD
results is still not very good but, with regard to all uncertainties, the differences seem to be tolerable.

We also achieve good agreement with the COMPASS data \ci{COMPASS-2019}
at $Q^2=2\,\gev^2$ and $x_B=0.093$, see lower right panel 
of Fig.\ \ref{fig:T}. Only our $t$-dependence
seems to be a bit flat. 
However, the COMPASS collaboration has a new, still preliminary set of data at the
same values of $t$ as in \ci{COMPASS-2019}. 
These new data, already shown at conferences \ci{peskova}, are
noticeably closer to our results.

\begin{figure}[th]
\begin{center}  
   \includegraphics[width=0.36\tw]{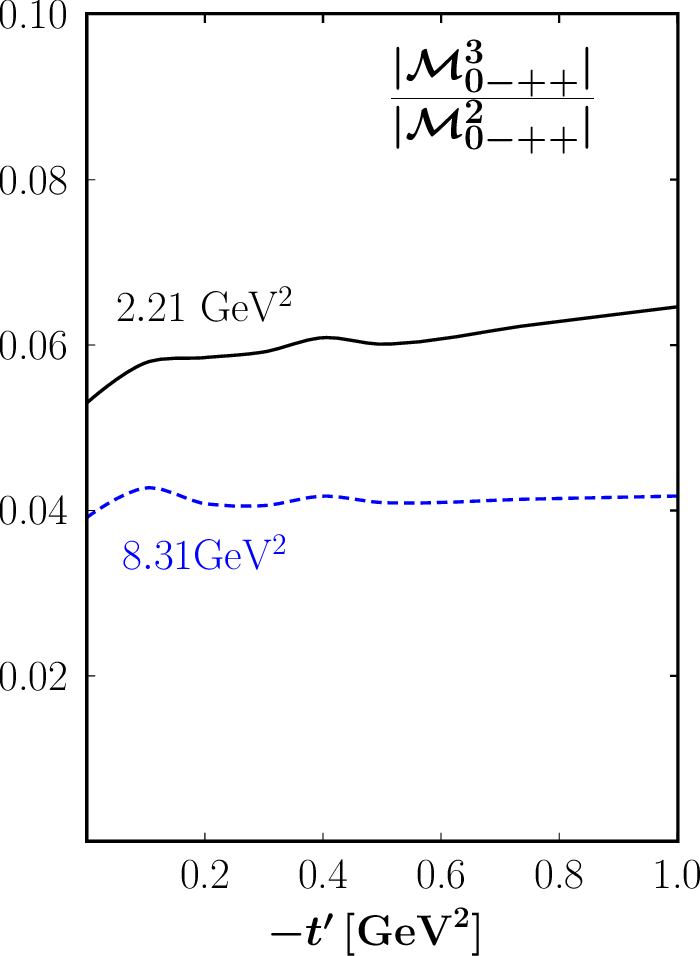}
  \end{center}
\caption{The ratio of 3-body and 2-body twist-3 contributions to the amplitude ${\cal M}_{0-,++}$ for
  two values of the photon virtuality.}
  \label{fig:compass}
\end{figure}

The 3-body contribution, i.e. the sum of the subprocess amplitudes \req{eq:sub-CF} and \req{eq:sub-CG} convoluted
with a transversity GPD, is much smaller than the 2-body, twist-3 one. This can be seen from 
Fig. \ref{fig:compass} where the ratio of the absolute values of these two contributions to the amplitude
${\cal M}_{0-,++}$ is displayed~\footnote{
         The 3-body contribution is given by a five dimensional integral which we evaluated with a Monte-Carlo
         procedure.}.
This ratio amounts only to about $5\%$, i.e. it is almost negligible. The corresponding ratio of these
contributions to the amplitude ${\cal M}_{0+,++}$ is of similar size. The smallness of the 3-body contribution
supports the assumption \req{eq:qbarqg-sudakov}.

We understand now why the results obtained with the DA \req{eq:3-body-da2} and the GPDs proposed in \ci{GK6,GK7,GK3}
are so close to those evaluated from the WW approximation in contrast to the KPK scenario. The 3-body contribution
is very small in both the scenarios but the DA $\phiPp$ generated from \req{eq:3-body-da2} through the
equation of motion is very close to unity except at extremely low scales, close to $\LQCD$, see Fig.\ \ref{fig:DA}.
The factor $f_{3\pi}\omega/(f_\pi\mu_\pi)$ in \req{eq:2-body-DA} is small for \req{eq:omega2}, about 0.08 at the
initial scale, while for the KPK scenario \req{eq:omega1} it is about -0.51.

\section{The collinear perturbative approach with massive gluons}
\label{sec:collinear}
A practical disadvantage of the MPA is the large computing time needed 
for cross sections evaluation, complicating large-scale fitting of
experimental data to extract the GPDs. 
A collinear approach is much faster since one has essentially
to evaluate only a two-dimensional integral,
while for the MPA three- and five-dimensional integrals contribute,
as discussed in Sec.\ \ref{sec:MPA}. 
Another disadvantage of the MPA is the demanding calculation 
of next-to-leading (NLO) corrections
due to the presence of $k_\perp$ terms.
On the other hand, the calculation of NLO corrections poses no principal difficulty 
for the collinear approach, although, at present, they have been calculated only 
for the twist-2 amplitude \ci{Belitsky:2001nq,Duplancic:2016bge}. 
However, in collinear approach the question then arises: 
how to regularize the end-point singularity appearing in the subprocess 
amplitude \req{eq:2-body-tw3-sub}?\footnote{
               In the photoproduction case ($\gamma q \to P q$)
                this contribution vanishes and no singularity appears.}.
A possibility is to use a dynamically generated gluon mass as a regulator. 
The idea of the gluon mass generation,
even if the local gauge symmetry of the QCD Lagrangian forbids a mass term, 
was proposed by Schwinger long ago in Ref. \cite{Schwinger:1962tn}, 
\cite{Schwinger:1962tp}. 
For a discussion of a dynamical generation of a gluon mass, 
see also \ci{Cornwall:1981zr}. The gluon mass generation, which is based on the 
Schwinger mechanism, is at present the subject of intensive studies 
in the QCD context and is motivated
by recent evidences for such a phenomenon from lattice simulations, for  reviews see e.g. \cite{Aguilar:2015bud},
\cite{Roberts:2021xnz}, \cite{Aguilar:2014tka}. 
Similar mass generation, also intensively studied at present,
can be achieved as a result of the formation of condensates \cite{Horak:2022aqx}, 
as well as within the instanton liquid model, 
see e.g. the review  \cite{Musakhanov:2021gof} and references therein.
The regularization of the end-point singularity can be thus achieved 
by introduction of a dynamically generated gluon mass into the gluon propagators. 
This  idea was applied in \cite{Radyushkin:2009zg} in a discussion of pion
electromagnetic form factor in perturbative QCD,
or recently in Ref. \cite{Shuryak:2020ktq} in a discussion of
the mesonic form factors~\footnote{
        In \cite{Shuryak:2020ktq} both quark transverse momentum and gluon mass 
        are suggested as regulators that originate from the subleading terms 
in denominators.}.

A comprehensive method for regularizing the end-point singularities 
with a dynamical gluon mass, $m_g$, would involve substituting it into all 
denominators of the gluonic propagators with momentum $k^\mu$. 
The substitution is made by 
replacing $k^2 + i \epsilon$ with $k^2 - m_g^2 + i \epsilon$. 
The integrations over the variables $x$ in the convolution with the GPDs 
and $\tau$ of the pion DAs are then performed.
However, in the present study we proceed in a simplified way which, we believe, 
permits to obtain a reliable estimate of the helicity amplitudes \req{eq:hel-amp}
easier. 
The minimal $m_g$ extension of the collinear approach 
inserts the gluon mass only in the gluon propagator that appears 
in the 2-body twist-3 subprocess amplitude \req{eq:2-body-tw3-sub}, 
the one in which the end-point singularity actually turns up.%
\footnote{This minimal extension is in agreement with the strategy 
outlined in Ref. \ci{anikin02}, 
where it was argued that for quantitative estimates, 
one should combine the factorizable
contributions and regularized non-factorizable ones.}
We think that this minimal extension is sufficient for our purpose.
The reason is that 
the twist-3 contribution regularized with a dynamical gluon mass 
differs substantially from the MPA with the WW approximation, 
as will become evident from our studies. Therefore, the GPDs derived in
\ci{GK6,GK7,GK3}
do not apply~\footnote{
As we already mentioned, the MPA takes effectively 
into account the transverse size of the meson. 
For deeply virtual Compton scattering, therefore, 
the collinear approach with the GPDs \ci{GK6,GK7,GK3}
derived from DVMP within the MPA should work, 
as is evidenced by \ci{sabatie}.}.
New extensive fitting of experimental data 
is required in order to obtain a new set of GPDs,
as well as possible inclusion of NLO corrections,
which is beyond the scope of the present paper.
With regard to all this, we will only present an exploratory
study of this approach
in order to demonstrate how the gluon mass regulates
the endpoint singularity in the 2-body twist-3 contribution.

  \subsection{ The gluon mass as a regulator to end-point singularities}
  \label{sec:gluon-mass}

The twist-3 subprocess amplitudes in the collinear limit are given 
in Sec.\ \ref{sec:tw3}. 
Here we explain the treatment of the double-poles and the end-point singularity 
in \req{eq:2-body-tw3-sub}.

\begin{figure}[t]
\begin{center}
\includegraphics[scale=0.6]{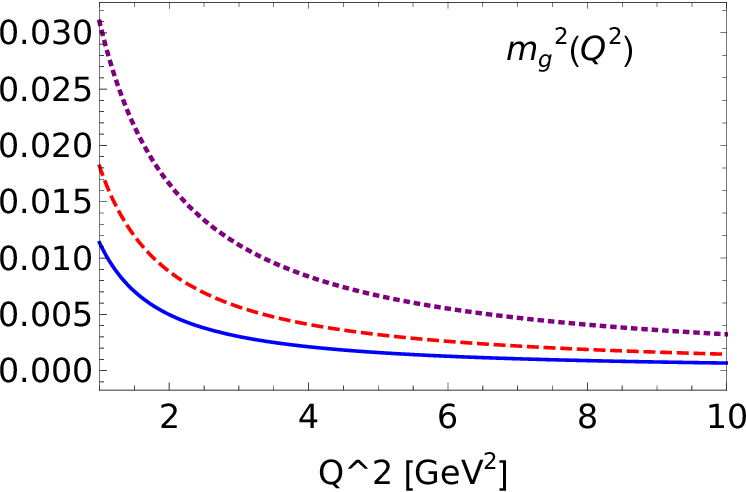}
\caption{The gluon mass $m_g^2(Q^2)$ as in \req{eq:mg2ev} for parameters
(381 MeV, 0.26), 
(436 MeV, 0.15), 
and
(557 MeV, 0.08),
denoted by solid, dashed and dotted line, respectively.
}
\label{fig:mg2ev} 
\end{center}
\end{figure}

We start with the regularization of the 2-body twist-3 subprocess amplitude \req{eq:2-body-tw3-sub} 
by the gluon mass. 
The double poles in the subprocess amplitude originate from a quark
and a gluon propagator. 
Changing the latter one by introducing the gluon mass the subprocess amplitude becomes
\ba
   {\cal H}^{\pi^i, q\bar{q}}_{0-\lambda,\mu\lambda}&=&\sqrt{2}\pi (2\lambda+\mu) \als(\mu_R) {\cal C}^{(ab)}_{\pi^i}
                             f_\pi\mu_\pi \frac{C_F}{N_C}\frac{Q^2}{\xi}   
                             \,\int_0^1 d\tau \phiPp(\tau) \nn\\
                         &&\times    \left[\frac{e_a}{(\sh+i\eps)(\taub \sh-m_g^2+i\eps)}
                           + \frac{e_b}{(\uh+i\eps)(\tau \uh-m_g^2+i\eps)}\right]
. \: \:
\label{eq:2-body-tw3-mg2}  
 \ea
Note that no double pole appears. 
Partial breaking of the propagator products as in \req{eq:partial}
(with $k_\perp^2$ being replaced by $m_g^2$) leads to a sum of single propagators. One can show that the
subprocess amplitude ${\cal H}^{\pi^i,q\bar{q}}$ is now regular. 
In this work we apply 
the scale dependent form 
\begin{equation}
m_g^2(Q^2)=\frac{m_0^2}{1+(Q^2/M^2)^{1+p}}
\, ,
\label{eq:mg2ev}
\end{equation}
for 
$m_0=376$ MeV 
and several parameter sets $(M,p)$
taken from 
a physically motivated fit
to the numerical solutions of the gluon
mass equation
given in \cite{Aguilar:2014tka} 
--
see Fig. \ref{fig:mg2ev}.

The 3-body twist-3 subprocess amplitudes 
\req{eq:3-body-tw3-CF} and \req{eq:3-body-tw3-CG} contain the double poles 
that do not cause any problem for integration over the GPDs that we are using. 
There, one may start from the distributions 
\be
\frac1{(\sh+i\eps)^2}\=-\frac{2\xi}{Q^2} \frac{d}{dx} \frac1{\sh+i\eps} \qquad
\frac1{(\uh+i\eps)^2}\=\frac{2\xi}{Q^2} \frac{d}{dx} \frac1{\uh+i\eps}
\, ,
\ee
which convoluted with a valence-quark transversity GPD, $K_T(x,\xi,t)$, lead to the form
\ba
\int_{-\xi}^1 dx \frac{K_T(x,\xi,t)}{(\sh+i\eps)^2} &=& \frac{2\xi}{Q^2} \int_{-\xi}^1dx \frac{K'_T(x,\xi,t)}{\sh+i\eps} \nn\\
&&- \,\frac{4\xi^2}{1-\xi} \frac1{Q^4} K_T(1,\xi,t) - \frac{2\xi}{Q^4} K_T(-\xi,\xi,t)
\, ,
\ea
and analogously for $1/\uh^2$. Since the GPDs are zero at $x=1$ and $x=-\xi$ (see the discussion in Sec. 
\ref{sec:GKgpds} and in particular Fig.\ \ref{fig:derivative}),  
the double pole convoluted with a GPD
reduces to a convolution of a single pole with the derivative of the GPD
\be
K'_T(x,\xi,t)\=\frac{d}{dx} K_T(x,\xi,t)\,.
\ee
The appearance of the GPD derivatives do not pose problems as long as they are continuous
at $x=\pm \xi$, which is the case for the parameterization of the GPDs we are using, see the discussion
in Sec.\ \ref{sec:GKgpds}.
The convolution of a valence-quark transversity GPDs and the double poles therefore reads
\ba
\lefteqn{\int_{-\xi}^1 dx K_T(x,\xi,t)\left[\frac{e_a}{(\sh+i\eps)^2} + \frac{e_b}{(\uh+i\eps)^2}\right]=}
                       \hspace*{0.1\tw} \nn\\
  &&\frac{4\xi^2}{Q^4} \int_{-\xi}^1 dx K'_T(x,\xi,t)\left[\frac{e_a}{x-\xi+i\eps} + \frac{e_b}{x+\xi-i\eps}\right]\,.
\label{eq:3-body-tw3-collCF}
\ea
The last term in \req{eq:3-body-tw3-CG} can be simplified with the help of identity
 \be
 \frac1{\sh\uh}\=-\frac1{Q^2}\left(\frac1{\sh} + \frac1{\uh}\right)\,.
 \ee
Thus, this is a single-pole contribution and numerically it turns out to be small. 
Since the $x$ and the $\tau$ integration 
factorize, 
one can perform the $\tau$ integrations
separately. Taking into account \req{eq:3-body-DA}, the $\tau$ integrals in 3-body contributions 
\req{eq:3-body-tw3-CF}  and \req{eq:3-body-tw3-CG}  amount to
\begin{equation}
\begin{array}{lclcl}
\omega_F=
\displaystyle
 \int_0^1\frac{d\tau}{\taub^2} 
\int_0^{\taub} \frac{d\tau_g}{\taub-\tau_g} \phiThreeP(\tau,\taub-\tau_g,\tau_g)
= 
\displaystyle
20 -\frac{15}{4} \omega_{1,0} +\frac{24}{5} \omega_{2,0} -\frac{6}{5} \omega_{1,1} 
\,,  \\[0.3cm]
\omega_G=
\displaystyle
\int_0^1\frac{d\tau}{\taub} 
\int_0^{\taub} \frac{d\tau_g}{\tau_g(\taub-\tau_g)} \phiThreeP(\tau,\taub-\tau_g,\tau_g)
=
\displaystyle
30
-10 \omega_{1,0}
+8 \omega_{2,0}
-\frac{1}{2} \omega_{1,1}
\, ,
\label{eq:intDA3pi}
\end{array}
\end{equation}
respectively.

\subsection{Results from the collinear approach}
\label{sec:coll-results}
\begin{figure}[h]
\begin{center}
\includegraphics[scale=0.7]{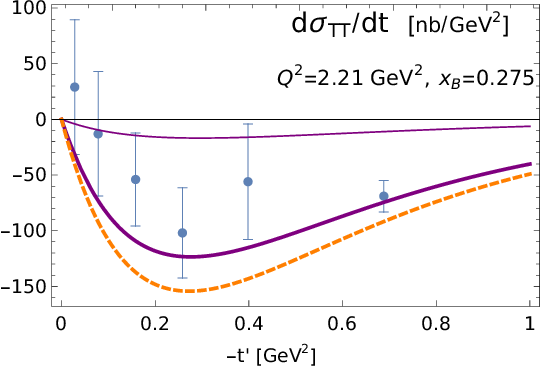}
\includegraphics[scale=0.7]{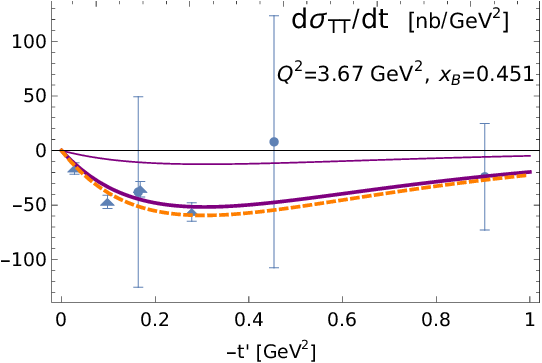}
\\
\includegraphics[scale=0.7]{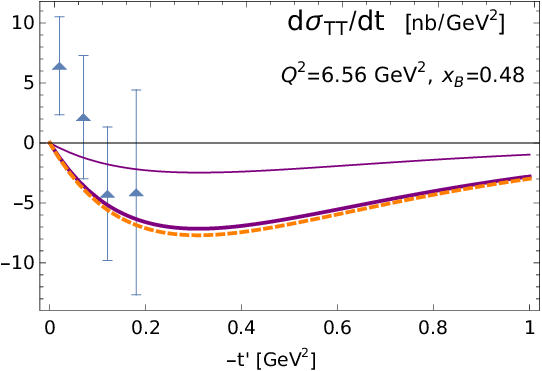}
\includegraphics[scale=0.7]{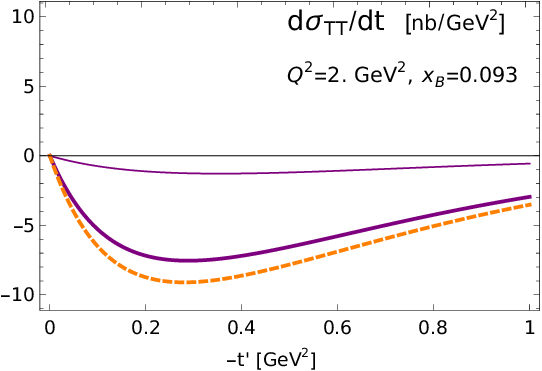}
\caption{The transverse-transverse interference
cross section 
\req{eq:sigmaTT} versus $t'$
for various kinematical settings,
obtained using 
the collinear approach with the gluon mass
\req{eq:mg2ev} for 
(436 MeV, 0.15).
The thick (thin) solid lines denote 
the cross sections obtained using the 
pion DA parameter sets 
  \req{eq:3-body-da2} 
  (\req{eq:3-body-da1}).
The dashed line represents the 
 WW prediction.
The experimental data are denoted as
filled circles \cite{CLAS14} and
triangles \cite{hall-A-2020}
(the triangles in the upper right plot
correspond to $Q^2=3.57\,\gev^2$ and $x_B=0.36$).
}
\label{fig:sigmaTT-coll} 
\end{center}
\end{figure} 

\begin{figure}[h!tb]
\begin{center}
\includegraphics[scale=0.7]{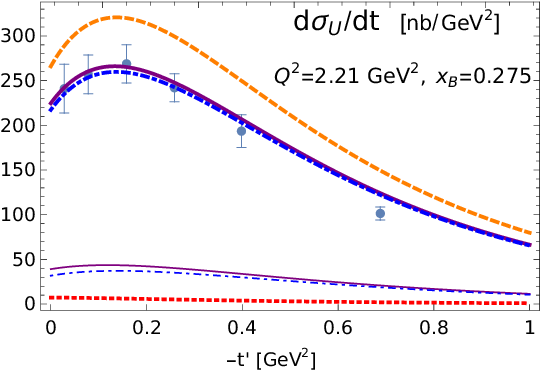}
\includegraphics[scale=0.7]{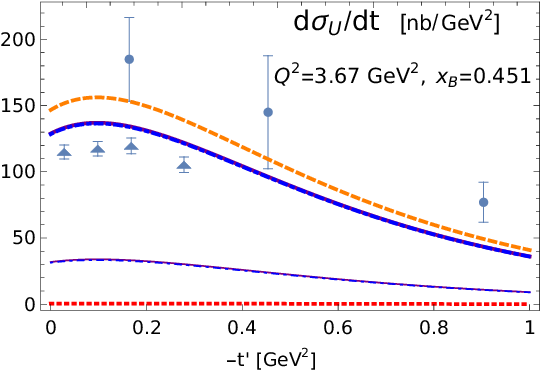}
\\
\includegraphics[scale=0.7]{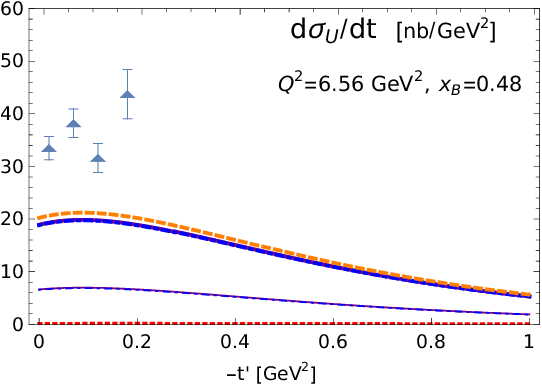}
\includegraphics[scale=0.7]{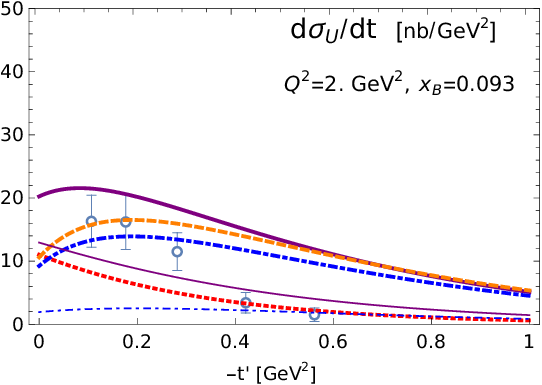}
\caption{The unseparated cross section 
\req{eq:sigU} versus $t'$
for various kinematical settings,
obtained using 
the collinear approach with the gluon mass
$m_g^2(Q^2)$ \req{eq:mg2ev} for 
(436 MeV, 0.15).
The thick (thin) lines denote 
the cross sections obtained using the 
pion DA parameter sets 
  \req{eq:3-body-da2} 
  (\req{eq:3-body-da1}):
solid line $d\sigma_U/dt$, 
dot-dashed line $d\sigma_T/dt$, 
and
dotted line $d\sigma_L/dt$. 
The dashed line represents the 
$d\sigma_U/dt$ WW contribution.
The open circles denote experimental data
\cite{COMPASS-2019} and for other notation 
we refer to Fig. 
\ref{fig:sigmaTT-coll}. 
}
\label{fig:sigmaU-coll} 
\end{center}
\end{figure}

We are now prepared to compute the subprocess amplitudes 
\req{eq:2-body-tw3-mg2}, \req{eq:3-body-tw3-CF}, 
and \req{eq:3-body-tw3-CG} within the collinear approach, 
followed by the determination of the s-channel helicity amplitudes 
\req{eq:hel-amp}, and ultimately, the transverse-transverse interference \req{eq:sigmaTT} 
($d\sigma_{TT}/dt$)
and unseparated \req{eq:sigU} 
($d \sigma_{U}/dt$)
cross sections. 
The expressions for computing the twist-2
contributions, and thus the longitudinal cross-section 
contributing to  $d \sigma_{U}/dt$,
are the usual ones \cite{bel-rad},
while  we use the standard twist-2 pion DA 
with the second Gegenbauer coefficient taken from \ci{braun15}.
Our current focus does not entail a comprehensive collinear analysis 
involving fits or the introduction of NLO corrections. 
Instead, we aim to present a proof of concept 
and offer 
an insight into the interplay of contributions. 
With this goal in mind, 
in order to simplify
our explorative study of collinear approach,
we 
employ an analytical integration of subprocess amplitudes 
over the GPD parameterization \cite{GK6,GK7,GK3}
and we omit the GPD evolution. 

Following the MPA analysis presented
in  Sec. \ref{sec:MPA-results},
on Figs.
\ref{fig:sigmaTT-coll} 
and
\ref{fig:sigmaU-coll} 
we compare our predictions
to a selected set of experimental results
from \cite{CLAS14}, \cite{hall-A-2020}
and \cite{COMPASS-2019}.
The thin and thick lines denote 
the cross sections obtained using the 
pion DA parameter sets 
\req{eq:3-body-da1} (KPK) 
and  \req{eq:3-body-da2}, while
the dashed lines represent the 
WW contributions in collinear approach.
The best description is 
obtained by using 
in \req{eq:mg2ev}
the parameters
$(M,p)=$(436 MeV, 0.15).

As in the MPA case, the predictions obtained 
using parameter set \req{eq:3-body-da2} 
are in good agreement with $d\sigma_{TT}/dt$ data
presented on Fig. \ref{fig:sigmaTT-coll}.
The agreement of the same set with $d\sigma_{U}/dt$ data
is good for the CLAS data at $Q^2=2.21$ GeV$^2$
and $x_B=0.275$, but fails for higher $Q^2$ and $x_B$.
Since only the GPD $\bar{E}_T$ contributes
to $d \sigma_{TT}/dt$, while both $H_T$ and
$\bar{E}_T$ contribute to
$d \sigma_{T}/dt$,
one can take it as a hint that $H_T$ needs to be
modified.
Similarly to the MPA case,
the collinear predictions obtained using the 
pion DA \req{eq:3-body-da1}
are too low.

\begin{figure}[t]
\begin{center}
\includegraphics[scale=0.5]{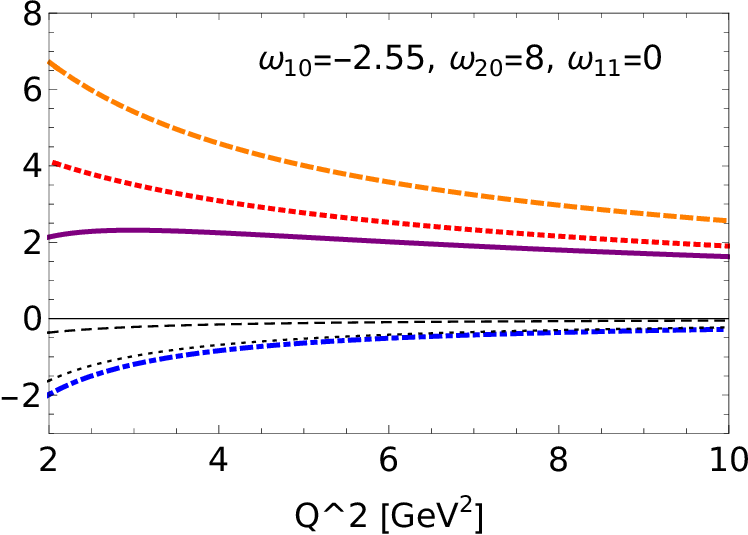}
\includegraphics[scale=0.5]{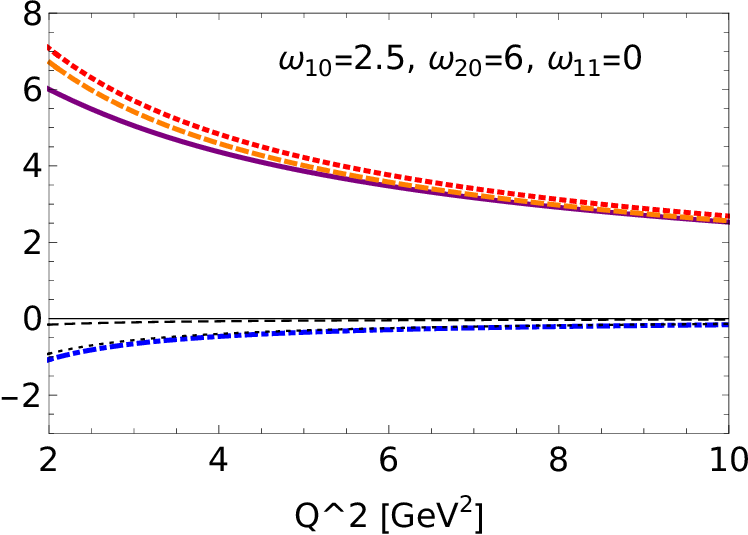}
\caption{
The sizes of pion DA contributions from
Eqs.  
\req{eq:2-body-tw3-sub}, \req{eq:3-body-tw3-CF} and \req{eq:3-body-tw3-CG}
and normalized as in \req{eq:DAcontr}
for $m_g^2(Q^2)$ \req{eq:mg2ev} and for parameters
(436 MeV, 0.15). The thick lines denote the WW (dashed), 2-body (dotted), 
3-body (dash-dotted), and the complete (solid) twist-3 relative contributions,
while the thin lines show the $C_F$ (dotted) and $C_G$ (dashed) proportional 3-body
twist-3 parts. 
Left plot is for pion DA set \req{eq:3-body-da1} 
and the right one for \req{eq:3-body-da2}.
}
\label{fig:DAcontr3} 
\end{center}
\end{figure}

From the series of plots in Fig. \ref{fig:sigmaU-coll}, 
it is apparent that the $Q^2$ and $x_B$ dependence 
of the $d \sigma_{U}/dt$ predictions obtained 
by set \req{eq:3-body-da2} does not seem to be satisfactory 
in the range considered. 
Notably, the decrease in collinear predictions 
obtained using the set \req{eq:3-body-da1} 
is much milder. 
As illustrated in Fig. \ref{fig:DAcontr3} 
(see Sec. 
\ref{sec:lessons}
for details),
this behavior is due to the interference 
of 2- and 3-body twist-3 contributions. 
This suggests investigating the direction 
in which the collinear approach
captures the observed $Q^2$ and $x_B$ dependence
with an appropriately modified pion DA.

On Fig. \ref{fig:sigmaU-coll} 
the longitudinal cross section $d\sigma_L/dt$
is depicted by a dashed line
and is 
much smaller than $d\sigma_T/dt$ 
for relatively low $Q^2$ and $W$ at which CLAS
and Hall A data are available.
In contrast, it is important to stress that 
in the low $x_B$ kinematics,
as for COMPASS data 
\cite{COMPASS-2019} depicted on Fig.
\ref{fig:sigmaU-coll} (bottom right figure),
the longitudinal cross section 
cannot be neglected and is of comparable
size as the transverse one.
Thus, in that energy region 
the NLO twist-2 corrections
should be included and 
the collinear approach 
is of particular importance
since it enables easier inclusion
of NLO corrections.
Analytical expressions for these corrections 
are available \ci{Belitsky:2001nq,Duplancic:2016bge}.
However, 
the twist-2 NLO results were not confronted with the data
but the model dependent assessments of the size of the NLO corrections
were given and they amount to $40-100\%$
\ci{Belitsky:2001nq,Diehl:2007hd}.
The calculation of NLO corrections to twist-3 part 
represent a demanding task left for future.
It is worth noting that if one introduces 
$m_g^2$ in gluon propagators for the twist-2 part, 
a new calculation even for NLO twist-2 corrections would be required. 
For higher $Q^2$, it's safe to neglect such $m_g^2/Q^2$ terms.
Since at $Q^2 \approx 2$ GeV$^2$ such corrections for LO 
contribute only up to a few percent, the corrections
to NLO are expected to be small as well.
The similar inclusion of $m_g^2$ terms in non-singular 3-body
twist-3 contributions could bring suppression of these terms
of up to $18\%$ for $Q^2=2$ GeV$^2$.
The suppression decreases fast with $Q^2$ and $x_B$.

\subsection{Lessons from DVMP and photoproduction}
\label{sec:lessons}
To better understand the obtained numerical results
it is instructive to analyze the relative sizes 
of the twist-3 contributions. 
We illustrate how the size of the twist-3 contribution
and its $Q^2$ dependence are influenced by the pion DA, 
and consequently by the interplay of the 2- and 3-body twist-3 contributions.
In order to do that we make an approximate factorization of $x$ and $\tau$ 
integration in \req{eq:2-body-tw3-mg2}, i.e, we use \req{eq:2-body-tw3-sub}
and regularize only the integral over $\tau$ by replacing   $\tau$ in the integrand of
\req{eq:2-body-tw3-sub} by  $\tau + \bm_g^2/Q^2$.
Consequently, the integral over  $\tau$ from \req{eq:2-body-tw3-sub} 
can be cast into the form
\begin{eqnarray}
\lefteqn{\int_0^1 d\tau 
\frac{1}{\tau + \bm_g^2/Q^2}
\, 
\phiPp(\tau)}
\nn \\ &=& 
\ln \left(1+\frac{1}{\bm_g^2/Q^2}\right) 
\left[
1+
 \frac{f_{3\pi}}{f_\pi\mu_\pi} \, \omega
	\left(1-\frac{30 \bm_g^4}{Q^4}-\frac{60 \bm_g^6}{Q^6}-\frac{30 \bm_g^8}{Q^8}\right) 
 \right]
	\nn \\ & &
	-
\frac{5}{2} 
 \frac{f_{3\pi}}{f_\pi\mu_\pi} \, \omega
 \left(1-\frac{4 \bm_g^2}{Q^2}-\frac{18 \bm_g^4}{Q^4}-\frac{12 \bm_g^6}{Q^6}\right) 
	\, ,
\label{eq:mg2int}
\end{eqnarray}
with $\omega$ given in \req{eq:omega}.
The first term 
in \req{eq:mg2int}
encapsulates the effect of end-point singularity.  
The values for 
integral \req{eq:mg2int}
decrease with increasing
$\bm_g^2$ and 
as expected vanish for
 $\bm_g^2 \to \infty $.
The justification for 
approximating 
\req{eq:2-body-tw3-mg2}
by 
\req{eq:2-body-tw3-sub}
supported by
\req{eq:mg2int}
lies 
in the smallness of the gluon mass.
For the small gluon mass we are employing, the numerical results 
are close to those obtained using \req{eq:2-body-tw3-mg2}.
This simplification facilitates a clearer interpretation 
of our numerical results from \req{eq:2-body-tw3-mg2}, 
allowing for a distinct separation of the roles played 
by GPDs (effectively overall factor) 
and the modifications introduced by the different 
choices for the twist-3 pion DA.

As an inspection of Eqs.\ \req{eq:2-body-tw3-sub},  
\req{eq:3-body-tw3-CF} and \req{eq:3-body-tw3-CG} 
reveals, the relative size of different twist-3 contributions
is essentially controlled by the integrals over the pion DAs%
\footnote{The contribution of remaining $C_G$ proportional term in \req{eq:3-body-tw3-CG}
is numerically small.},
i.e., the size of 2-body twist-3 contribution is proportional to
\begin{subequations}
\label{eq:DAcontr}
\begin{equation}
\frac{\mu_\pi}{Q^2}
\,
C_F
\left\{
\ln \left(1+\frac{1}{\bm_g^2/Q^2}\right) 
+
 \frac{f_{3\pi}}{f_\pi\mu_\pi} \, \omega
\left[
\ln \left(1+\frac{1}{\bm_g^2/Q^2}\right) 
-
\frac{5}{2} 
+
{\cal O}\left(\frac{\bm_g^2}{Q^2}\right) 
 \right]
 \right\}
\, ,
\label{eq:DAcontr-2b}
\end{equation}
while 
the $C_F$ and $C_G$ proportional 
3-body twist-3 contributions
are governed by 
\begin{eqnarray}
-\frac{\mu_\pi}{Q^2}
\,
C_F \,\frac{f_{3\pi}}{f_\pi\mu_\pi} \, 
\omega_F
\, 
,
& &
\frac{\mu_\pi}{Q^2}
\, 
C_G \,\frac{f_{3\pi}}{f_\pi\mu_\pi} \, 
\, 
\,
\omega_G
\, ,
\label{eq:DAcontr-3b}
\end{eqnarray}
\end{subequations}
respectively.
The factors $\omega$, $\omega_F$ and $\omega_G$
depend on pion DA coefficients
$\omega_{1,0}$,
$\omega_{2,0}$,
and
$\omega_{1,1}$,
and they are defined in
\req{eq:omega}
and
\req{eq:intDA3pi}.
The WW approximation corresponds to taking 
$\omega=\omega_F=\omega_G=0$.

Now it is straightforward to illustrate the sizes
of different twist-3 contributions for 
selected pion DAs.
In Fig. \ref{fig:DAcontr3}
we compare the sizes of 
the contributions \req{eq:DAcontr}
in the $2\leq Q^2 \leq 10$ GeV$^2$ range.
The gluon mass $m_g(Q^2)$ is 
introduced as in \req{eq:mg2ev}.
We compare
both DA parameter sets introduced above, 
i.e., 
\req{eq:3-body-da1} and 
\req{eq:3-body-da2},  
with their evolution taken into account.
The corresponding values for $\omega$ are given
in \req{eq:omega1}
and \req{eq:omega2},
while
\begin{equation}
\omega_F(\mu_0^2)=67.96
\, ,
\quad
\quad
\frac{C_G}{C_F}\omega_G(\mu_0^2)=-14.94
\, ,
\end{equation}
and
\begin{equation}
\omega_F(\mu_0^2)=39.43
\, ,
\quad
\quad
\frac{C_G}{C_F}\omega_G(\mu_0^2)=-6.63
\, ,
\end{equation}
respectively.
To remind,
the $\pi^0$ photoproduction \cite{KPK21} was used for the
determination of the parameter set \req{eq:3-body-da1}. 
The parameter set \req{eq:3-body-da2} was introduced 
in Sec. \ref{sec:DA} in order to reconcile
the DVMP and photoproduction data
along with minimal adjustments to the GPD parameters 
given in Tab. \ref{tab:GPD-parameters}. 

The dominant $C_F$ proportional 3-body twist-3 contribution
in DVMP
is proportional to $\omega_F$, defined by the integral in \req{eq:intDA3pi}.
This integral, and consequently  $\omega_F$, 
also appears in and dominates the $\pi^0$ photoproduction.
Additionally, since the twist-3 contribution proportional to $\phiPp$ vanishes 
in photoproduction, the twist-3 contribution proportional to $\phiThreeP$ governs, 
establishing the range of $\omega_F$ through photoproduction.

Although the prefactor 
$f_{3\pi}/(f_\pi\mu_\pi)$ 
from \req{eq:DAcontr}
is a small number 
($0.01515$ at the initial scale $\mu_0=2 \gev$ and it decreases), 
the effect of 3-body pion DA encapsulated in $\omega$ proportional
terms can significantly alter the 2-body twist-3 contribution, 
especially at lower $Q^2$ values.
This is prominent for the set \req{eq:3-body-da1}, 
where $\omega$ takes a large negative value \req{eq:omega1}.
In the left Fig. \ref{fig:DAcontr3} the 2-body twist-3 contribution
lies well beyond the WW prediction.
The 3-body twist-3 contribution is negative and large
for lower $Q^2$ values, making our twist-3 prediction
much lower than the WW prediction. 
Since in the MPA the WW predictions along with GPD
parameters quoted in Tab. \ref{tab:GPD-parameters} describe the data well,
the set \req{eq:3-body-da1} 
produces the predictions in both the MPA
and the collinear approach that do not match the data.
This could also be seen as a hint to modify the GPDs but 
the constraints on their form coming from other sources
(as, for example, lattice QCD) do not leave enough room to
incorporate this particular set.
But it is important to note that for the set \req{eq:3-body-da1} 
the cancellation between 2- and 3-body contributions 
results in an extremely mild dependence on $Q^2$, 
which, as discussed above, aligns with the $Q^2$ dependence of the data.
Hence, to accurately capture the $Q^2$ dependence of the data,
a corresponding set may be constructed in a similar manner,
using the 3-body twist-3 contributions to modify the steep decent
of the 2-body twist-3 contributions.

For the set \req{eq:3-body-da2}, with a small positive value \req{eq:omega2}, 
the 2-body twist-3 contribution closely aligns with the WW prediction. 
The 3-body twist-3 contribution is negative, and since its dominant part 
is also the major contributor to the photoproduction, 
it is smaller than the 2-body contribution but not negligible. 
We note that the incorporation of the gluon mass into the 3-body twist-3 component 
would result in a further decrease in this contribution, 
similar to the observed effects in the MPA.
By design, this set effectively describes the DVMP data within the MPA, 
and its predictions do not differ significantly from the WW prediction
both in the MPA and the collinear approach. 

The above analysis illustrates the potential to modify the pion DA expansion 
coefficients by considering both DVMP and photoproduction data.
It 
aims to enhance our understanding of the numerical results 
obtained through both the collinear approach and the MPA, 
and should serve as a guide for the use of the collinear 
approach in future fits. As in other DVMP scenarios, there are three 
potential approaches: using the meson DA from another process/input 
and fitting the GPDs, retaining the GPDs and attempting to fit the meson DA, 
or trying to fit both simultaneously. 
However, the quality of the data may pose challenges in effectively 
fitting both DA and GPDs.  In this study we refrained 
from attempting the fits, reserving them for future work. 

\section{Summary}
\label{sec:summary}
We studied the twist-3 contributions to DVMP beyond the WW approximation. 
The 3-body twist-3 DA is fixed by 
the adjustment to
the wide-angle pion photoproduction data. This DA generates modifications of
the flat 2-body twist-3 DA, $\phiPp\equiv 1$, through the equation of motion. 
Still, the new DA $\phiPp$ does not vanish at the end-points, $\tau=0$ and $1$. 
As in \ci{GK5,GK6}, we apply the MPA, 
in which quark transverse momenta and Sudakov suppressions 
are taken into account, in order
to regularize the end-point singularity present in the 2-body twist-3 contribution. 
The modifications of the twist-3 subprocess amplitude are small
so that, within the MPA, the GPDs derived in \ci{GK5,GK6} 
(with only minor adjustments applied)
still lead to reasonable agreement with experiment. 
As a second regularization method, we proposed the use of a dynamically generated gluon mass 
in combination with the collinear approach. 
The agreement with the experiment is only fair due to the fact that 
the soft parameters were left the same as in the MPA case. 
The interplay between different contributions has been illustrated, 
and a basis for a more thorough analysis, 
including higher-order corrections, has been outlined.
It was found that NLO twist-2 corrections could play an important role 
in COMPASS kinematics.

We stress that the twist-3 analysis connects deeply virtual processes
(probing the GPDs at small $-t$) with wide-angle ones (probing GPDs at large $-t$)
and allows us to extract information on the GPDs 
at a fairly large range of $t$. 
This is valuable information required for the study of the 3-dimensional
partonic structure of the proton.\\

{\it Acknowledgments} 
This publication is supported 
by the Croatian Science Foundation project IP-2019-04-9709,
and by the EU Horizon 2020 research and innovation programme, STRONG-2020
project, under grant agreement No 824093.
The work of L. S. is supported by the grant 2019/33/B/ST2/02588 
of the National Science Center in Poland. 
L. S. thanks the P2IO Laboratory of Excellence 
(Programme Investissements d’Avenir ANR-10-LABEX-0038) 
and the P2I - Graduate School of Physics of Paris-Saclay University 
for support.



\begin{thebibliography}{99}
  \label{sec:biblio}

\bibitem{collins96} J.~C.~Collins, L.~Frankfurt and M.~Strikman,
Phys. Rev. D \textbf{56}, 2982-3006 (1997)
[arXiv:hep-ph/9611433 [hep-ph]].

\bibitem{defurne} M.~Defurne \textit{et al.} [Jefferson Lab Hall A],
Phys. Rev. Lett. \textbf{117}, no.26, 262001 (2016)
[arXiv:1608.01003 [hep-ex]].

\bibitem{mazouz}M.~Mazouz \textit{et al.} [Jefferson Lab Hall A],
Phys. Rev. Lett. \textbf{118}, no.22, 222002 (2017)
[arXiv:1702.00835 [hep-ex]].

\bibitem{hermes10} A.~Airapetian \textit{et al.} [HERMES],
Phys. Lett. B \textbf{682}, 345-350 (2010)
[arXiv:0907.2596 [hep-ex]].

\bibitem{CLAS14} I.~Bedlinskiy \textit{et al.} [CLAS],
Phys. Rev. C \textbf{90}, no.2, 025205 (2014)
[arXiv:1405.0988 [nucl-ex]].

\bibitem{GK5} S.~V.~Goloskokov and P.~Kroll,
Eur. Phys. J. C \textbf{65}, 137-151 (2010)
[arXiv:0906.0460 [hep-ph]].

\bibitem{frankfurt} L.~Frankfurt, W.~Koepf and M.~Strikman,
Phys. Rev. D \textbf{54}, 3194-3215 (1996)
[arXiv:hep-ph/9509311 [hep-ph]].

\bibitem{GK6} S.~V.~Goloskokov and P.~Kroll,
Eur. Phys. J. A \textbf{47} (2011), 112
[arXiv:1106.4897 [hep-ph]]. 

\bibitem{PK-kaon} P.~Kroll,
Eur. Phys. J. A \textbf{55}, no.5, 76 (2019)
[arXiv:1901.11380 [hep-ph]].

\bibitem{goldstein}G.~R.~Goldstein, J.~O.~Gonzalez Hernandez and S.~Liuti,
Phys. Rev. D \textbf{91}, no.11, 114013 (2015)
[arXiv:1311.0483 [hep-ph]].


\bibitem{GK7} S.~V.~Goloskokov and P.~Kroll,
Eur. Phys. J. C \textbf{74}, 2725 (2014)
[arXiv:1310.1472 [hep-ph]].

\bibitem{hermes-SDME} A.~Airapetian \textit{et al.} [HERMES],
Eur. Phys. J. C \textbf{62}, 659-695 (2009)
[arXiv:0901.0701 [hep-ex]].

\bibitem{compass-SDME}  G.~D.~Alexeev \textit{et al.} [COMPASS],
Eur. Phys. J. C \textbf{83}, no.10, 924 (2023)
[arXiv:2210.16932 [hep-ex]].

\bibitem{anikin02}
I.~V.~Anikin and O.~V.~Teryaev,
Phys. Lett. B \textbf{554}, 51-63 (2003)
[arXiv:hep-ph/0211028 [hep-ph]].


\bibitem{KPK21} P.~Kroll and K.~Passek-Kumeri\v{c}ki,
Phys. Rev. D \textbf{104}, no.5, 054040 (2021)
[arXiv:2107.04544 [hep-ph]].

\bibitem{Belitsky:2000vx}
A.~V.~Belitsky and D.~Mueller,
Nucl. Phys. B \textbf{589}, 611-630 (2000)
[arXiv:hep-ph/0007031 [hep-ph]].

\bibitem{ji98} X.~D.~Ji,
J. Phys. G \textbf{24}, 1181-1205 (1998)
[arXiv:hep-ph/9807358 [hep-ph]].

\bibitem{braun14} V.~M.~Braun, A.~N.~Manashov, D.~M\"uller and B.~M.~Pirnay,
Phys. Rev. D \textbf{89}, no.7, 074022 (2014)
[arXiv:1401.7621 [hep-ph]].

\bibitem{GK3} S.~V.~Goloskokov and P.~Kroll,
Eur. Phys. J. C \textbf{53}, 367-384 (2008)
[arXiv:0708.3569 [hep-ph]].

\bibitem{musatov99} I.~V.~Musatov and A.~V.~Radyushkin,
Phys. Rev. D \textbf{61}, 074027 (2000)
[arXiv:hep-ph/9905376 [hep-ph]].

\bibitem{DFJK4} M.~Diehl, T.~Feldmann, R.~Jakob and P.~Kroll,
Eur. Phys. J. C \textbf{39}, 1-39 (2005)
[arXiv:hep-ph/0408173 [hep-ph]].

\bibitem{teramond} G.~F.~de Teramond \textit{et al.} [HLFHS],
Phys. Rev. Lett. \textbf{120}, no.18, 182001 (2018)
[arXiv:1801.09154 [hep-ph]].

\bibitem{ABM12} S.~Alekhin, J.~Blumlein and S.~Moch,
Phys. Rev. D \textbf{86}, 054009 (2012)
[arXiv:1202.2281 [hep-ph]].

\bibitem{FSSV09} D.~de Florian, R.~Sassot, M.~Stratmann and W.~Vogelsang,
Phys. Rev. D \textbf{80}, 034030 (2009)
[arXiv:0904.3821 [hep-ph]].

\bibitem{diehl03} M.~Diehl,
Phys. Rept. \textbf{388}, 41-277 (2003)
[arXiv:hep-ph/0307382 [hep-ph]].

\bibitem{kivel} N.~Kivel, M.~V.~Polyakov, A.~Schafer and O.~V.~Teryaev,
Phys. Lett. B \textbf{497}, 73-79 (2001)
[arXiv:hep-ph/0007315 [hep-ph]].
  
\bibitem{aslan} F.~Aslan, M.~Burkardt, C.~Lorc\'e, A.~Metz and B.~Pasquini,
Phys. Rev. D \textbf{98}, no.1, 014038 (2018)
[arXiv:1802.06243 [hep-ph]].

\bibitem{bel-rad} A.~V.~Belitsky and A.~V.~Radyushkin,
Phys. Rept. \textbf{418}, 1-387 (2005)
[arXiv:hep-ph/0504030 [hep-ph]].

\bibitem{LQCD} M.~Gockeler \textit{et al.} [QCDSF and UKQCD],
Phys. Lett. B \textbf{627}, 113-123 (2005)
[arXiv:hep-lat/0507001 [hep-lat]].

\bibitem{LQCD2} 
M.~G\"ockeler \textit{et al.} [QCDSF and UKQCD],
Phys. Rev. Lett. \textbf{98}, 222001 (2007)
[arXiv:hep-lat/0612032 [hep-lat]].

\bibitem{bertone21}V.~Bertone, H.~Dutrieux, C.~Mezrag, H.~Moutarde and P.~Sznajder,
Phys. Rev. D \textbf{103}, no.11, 114019 (2021)
[arXiv:2104.03836 [hep-ph]].

\bibitem{braun-filyanov} V.~M.~Braun and I.~E.~Filyanov,
Z. Phys. C \textbf{48}, 239-248 (1990).

\bibitem{KPK18} P.~Kroll and K.~Passek-Kumeri\v{c}ki,
Phys. Rev. D \textbf{97} (2018) no.7, 074023
[arXiv:1802.06597 [hep-ph]].

\bibitem{ball} P.~Ball,
JHEP \textbf{01}, 010 (1999)
[arXiv:hep-ph/9812375 [hep-ph]].

\bibitem{kunkel17} M.~C.~Kunkel \textit{et al.} [CLAS],
Phys. Rev. C \textbf{98}, no.1, 015207 (2018)
[arXiv:1712.10314 [hep-ex]].


\bibitem{sotiropoulos}  M.~G.~Sotiropoulos and G.~F.~Sterman,
Nucl. Phys. B \textbf{425}, 489-515 (1994)
[arXiv:hep-ph/9401237 [hep-ph]].

\bibitem{bolz} J.~Bolz, R.~Jakob, P.~Kroll, M.~Bergmann and N.~G.~Stefanis,
Z. Phys. C \textbf{66}, 267-278 (1995)
[arXiv:hep-ph/9405340 [hep-ph]].

\bibitem{botts89} J.~Botts and G.~F.~Sterman,
Nucl. Phys. B \textbf{325}, 62-100 (1989).


\bibitem{hall-A-2020} M.~Dlamini \textit{et al.} [Jefferson Lab Hall A],
Phys. Rev. Lett. \textbf{127}, no.15, 152301 (2021)
[arXiv:2011.11125 [hep-ex]].

\bibitem{COMPASS-2019} M.~G.~Alexeev \textit{et al.} [COMPASS],
Phys. Lett. B \textbf{805}, 135454 (2020)
[arXiv:1903.12030 [hep-ex]].

\bibitem{peskova} M.~Peskova, K.~Lavickova and N.~d'Hose, on behalf of the COMPASS collaboration,
  PoS ICHEP2022, 832 (2022). 

\bibitem{Belitsky:2001nq}
A.~V.~Belitsky and D.~Mueller,
Phys. Lett. B \textbf{513} (2001), 349-360
[arXiv:hep-ph/0105046 [hep-ph]].

\bibitem{Duplancic:2016bge}
G.~Duplan\v{c}i\'c, D.~M\"uller and K.~Passek-Kumeri\v{c}ki,
Phys. Lett. B \textbf{771} (2017), 603-610
[arXiv:1612.01937 [hep-ph]].

\bibitem{Schwinger:1962tn}
J.~S.~Schwinger,
Phys. Rev. \textbf{125} (1962), 397-398

\bibitem{Schwinger:1962tp}
J.~S.~Schwinger,
Phys. Rev. \textbf{128} (1962), 2425-2429

\bibitem{Cornwall:1981zr} J.~M.~Cornwall,
Phys. Rev. D \textbf{26} (1982), 1453

\bibitem{Aguilar:2015bud}
A.~C.~Aguilar, D.~Binosi and J.~Papavassiliou,
Front. Phys. (Beijing) \textbf{11} (2016) no.2, 111203
[arXiv:1511.08361 [hep-ph]].

\bibitem{Roberts:2021xnz}
C.~D.~Roberts,
AAPPS Bull. \textbf{31} (2021), 6
[arXiv:2101.08340 [hep-ph]].

\bibitem{Aguilar:2014tka}
A.~C.~Aguilar, D.~Binosi and J.~Papavassiliou,
Phys. Rev. D \textbf{89} (2014) no.8, 085032
[arXiv:1401.3631 [hep-ph]].

\bibitem{Horak:2022aqx}
J.~Horak, F.~Ihssen, J.~Papavassiliou, J.~M.~Pawlowski, A.~Weber and C.~Wetterich,
SciPost Phys. \textbf{13} (2022) no.2, 042
[arXiv:2201.09747 [hep-ph]].

\bibitem{Musakhanov:2021gof}
M.~Musakhanov and U.~Yakhshiev,
Int. J. Mod. Phys. E \textbf{30} (2021) no.11, 2141005
[arXiv:2103.16628 [hep-ph]].

\bibitem{Radyushkin:2009zg}
A.~V.~Radyushkin,
Phys. Rev. D \textbf{80} (2009), 094009
[arXiv:0906.0323 [hep-ph]].

\bibitem{Shuryak:2020ktq}
E.~Shuryak and I.~Zahed,
Phys. Rev. D \textbf{103} (2021) no.5, 054028
[arXiv:2008.06169 [hep-ph]].

\bibitem{sabatie}P.~Kroll, H.~Moutarde and F.~Sabatie,
Eur. Phys. J. C \textbf{73}, no.1, 2278 (2013)
[arXiv:1210.6975 [hep-ph]].

\bibitem{braun15} V.~M.~Braun, S.~Collins, M.~Göckeler, P.~Pérez-Rubio, A.~Sch\"{a}fer, 
  R.~W.~Schiel and A.~Sternbeck,
  Phys.\ Rev.\ D {\bf 92}, no. 1, 014504 (2015)
  [arXiv:1503.03656 [hep-lat]].

\bibitem{Diehl:2007hd}
M.~Diehl and W.~Kugler,
Eur. Phys. J. C \textbf{52} (2007), 933-966
[arXiv:0708.1121 [hep-ph]].


\end{thebibliography}
\end{document}